\documentstyle[prd,aps,eqsecnum,array,epsf]{revtex}

\begin{document}

\title{Affinity for Scalar Fields to Dissipate}

\author{Arjun Berera$^{1}\;$\thanks{E-mail:
ab@ph.ed.ac.uk; PPARC Advanced Fellow}$\;$,
Rudnei O. Ramos$^{2,3}\;
$\thanks{E-mail: rudnei@peterpan.dartmouth.edu.
$^{3}\;$Permanent address.}}

\address{
{\it $^{1}\;$ Department of Physics and Astronomy, University of 
Edinburgh,}\\
{\it Edinburgh, EH9 3JZ, United Kingdom}
\\
{\it $^{2}\;$ Department of Physics and Astronomy, Dartmouth College,}\\
{\it  Hanover, New Hampshire 03755-3528, USA}\\
{\it and $^{3}$Departamento de F\'{\i}sica Te\'orica,}
{\it Instituto de F\'{\i}sica, Universidade do Estado do Rio de Janeiro,}\\
{\it 20550-013 Rio de Janeiro, RJ, Brazil}}

\maketitle

\begin{abstract}

The zero temperature  effective equation of motion 
is derived for a scalar field
interacting with other fields.
{}For a broad range of cases, involving interaction
with as few as one or two fields, dissipative regimes are found
for the scalar field system.
The zero temperature limit constitutes a baseline effect that
will be prevalent in any general statistical state.
Thus, the results found here provide strong evidence that dissipation
is the norm not exception for an interacting scalar field system.
{}For application to inflationary cosmology, this provides
convincing evidence that warm
inflation could be a natural dynamics once proper treatment of
interactions is done.  The results found here also may have applicability
to entropy production during the chiral phase transition in
heavy ion collision.

\vspace{0.34cm}
\noindent
PACS number(s): 98.80 Cq, 05.70.Ln, 11.10.Wx

\end{abstract}

\vspace{0.5cm}

\centerline{\large \bf In Press Physical Review D (2001)}
\centerline{\bf hep-ph/0101049}

\section{Introduction}
\label{sect1}

Statistical mechanics generally expects an interacting system
to equally distribute its energy amongst all the constituent
degrees of freedom.  Efforts to understand this most basic statement
from quantum field theory have focused on simple dissipative
systems.  The most common example amongst these is the case
of a quantum scalar field which is interacting with
other fields and which has an amplitude that is initially
displaced from equilibrium. 
There has been considerable interest in this kind of problem
due to its many applications, ranging from
the dynamics of condensates in condensed matter physics to the 
evolution of a inflaton field in inflationary cosmology
(for a recent review see \cite{rev}).
Several methods have been applied to this problem
involving analytic and numerical analysis
\cite{boya,boya1,boya3,habib,hu3,baacke2,hs,morikawa,ian,GR,BGR,hu,hu4}
as well as lattice simulations \cite{patkos,wett}.

Among the variety of dissipative dynamics
that can be studied for such scalar field systems,
the most tractable, given the
present state of understanding in quantum statistical
mechanics, is the adiabatic regime, where the motion
of the scalar field amplitude is slow.
One of the first implementation of this approximation was due
to Caldeira and Leggett \cite{cl} for the problem of interacting
quantum mechanical harmonic oscillators. Since then,
generalizations of this kind of dynamics for the case of scalar 
fields have been implemented by several authors 
\cite{hs,morikawa,ian,GR,BGR}.
In particular, in an
earlier work by us \cite{BGR}, a specific consistent 
solution regime for adiabatic dissipative dynamics was
identified, in which the motion of the scalar field
amplitude is overdamped. Our treatment in that work was
restricted to the high-temperature regime.  The purpose of this paper
is to extend those results to zero temperature and to find valid
parameter regions for the adiabatic dissipative regime.
The key step is the inclusion of quasi-particles effects 
through the appropriate use of full two-point Green's functions
at zero temperature.  This plays a central role in the dynamics,
as will be reviewed later in this paper.

A primary, but not singular, motivation for understanding the overdamped 
regime is to realize the warm inflation scenarios from
first principles.  The extension of the overdamped solution
to zero temperature has important consequences for this
goal.  A careful examination of the earlier work \cite{BGR}
as well as \cite{BGR2,abadiab}  reveal that the high-temperature
overdamped solutions came very close to a full
realization of warm inflation from simple interacting models.
Although ultimately what we found for that regime
was that full warm inflation
solutions only could be constructed for certain
complicated models \cite{BGR2,abadiab,linde},
a foremost impediment to simpler solutions was the limitations
of the high temperature approximation.  The
main accomplishment of this paper is to
establish that the extension of the solution to $T=0$
can be adequately posed.  Although only flat nonexpanding
spacetime is treated in this paper, these results provide
a necessary step towards the case of expanding spacetime.
{}Furthermore, the results in this paper already are suggestive that
once extension to expanding spacetime
is done, warm inflation solution probably will be viable in most of the 
simple quantum field theory models.

The calculation in this paper is incomplete in that
dissipation at zero temperature necessarily implies a
nonequilibrium state, which immediately should be driven up
to some excited statistical state.  As such, the results presented here
serve only as an indication for a nontrivial
dissipative effect, which requires a more general nonequilibrium
treatment for determining the precise nature of
the excited statistical state.  In this respect, the
results in this paper are as yet insufficient for
constructing completely consistent quantum field theory warm inflation 
solutions. 
Nevertheless, these calculations establish an important point,
that interacting scalar field systems have an intrinsic
and, in appropriate regimes, robust affinity to dissipate their energy.
This fact in turn could place considerable doubt
on a commonly followed, though unproven,
assumption of inflationary cosmology, that dissipative effects
of the scalar inflaton field can be ignored, thereby
leading to a supercooled regime of inflation with vanishing, or 
negligible production of radiation.

Despite the unconventional implications our results
suggest for inflationary cosmology, the argument for
supercooling in scalar field driven inflation
deserves a reexamination for the following
reason.  The basic statement of supercooled
inflation is that a single degree of freedom,
the zero mode of the inflaton, maintains all the energy
of the universe during the entire duration of
inflation in the form of potential,
or equivalently vacuum, energy.  However,  the curious 
point is, even if this
single degree were to allow a minuscule fraction
of the energy to be released, say one part in $10^{20}$,
it still would constitute a significant radiation
energy density component in the universe.
{}For example, for inflation with vacuum energy
at the GUT scale $\sim 10^{15-16} {\rm GeV}$,
leaking one part in $10^{20}$ of this energy into radiation 
corresponds to a temperature of $10^{11} {\rm GeV}$,
which is nonnegligible.  In fact, the most relevant
lower bound that cosmology places on the temperature after inflation
comes from the success of hot Big-Bang nucleosynthesis,
which thus requires the universe to be within the
radiation dominated regime by $T \gtrsim 1 {\rm GeV}$.
This limit can be met in the above example by dissipating 
as little as one part in $10^{60}$ of the vacuum energy
into radiation.
Thus, from the perspective of both interacting field theory and
basic notions of equipartition,  it appears to be a highly
tuned requirement of supercooled inflation to prohibit
the inflaton from even such tiny amounts of dissipation.
On the flip side, the warm inflation picture demonstrates
\cite{wi} that by relaxing this requirement,
the most unnatural and technically intractable aspect of supercooled
inflation, reheating, becomes unnecessary.
As such, despite the technical complications in computing dissipative
effects, as exemplified by this and the earlier papers,
on general grounds
their presence should be a natural expectation.

The paper is organized as follows.  Our real-time dissipative
formalism is reviewed in Sec. \ref{sect2}.  Explicit expressions
are given here for the real-time, 
fully dressed Bose and Fermi two-point Green's
functions, which are valid in the entire temperature range,
including $T=0$.  This formalism then is applied
in Sects. \ref{sect2}  and \ref{sect3} to obtain the
effective equation of motion of the system, here a scalar field that
has a classical amplitude out
of equilibrium at $T=0$, which is coupled to fields
that act as a reservoir bath for dissipative energy exchange.
Two types of models are treated
in Sections \ref{sect2} and \ref{sect3}, which
we have denoted respectively as direct decay models, where the 
system field directly decays to light particles that are part of 
the environment, and indirect decay models, 
where this process is mediated by intermediate fields.
In Sec. \ref{sect4}, we apply the effective equations of motion
to estimate the magnitude of radiation energy production.
{}Finally in Sec. \ref{sect5}, we give concluding remarks and
comment on the implication of our results for the case
of an expanding spacetime, which is relevant for
inflationary cosmology.  An Appendix is also included 
in which the 
computations are given of the 
zero temperature decay widths that are used in the paper.

\section{Direct Decay Models}
\label{sect2}

In this section, models will be studied where the system,
here a scalar field $\Phi$, is coupled to bath fields
of lighter mass than itself, thus allowing $\Phi$ to
directly decay into the bath fields.
In the first model the scalar field interacts
with a set of $N_\psi$ fermion fields with Lagrangian density

\begin{eqnarray}
{\cal L} [ \Phi, \bar{\psi}_k, \psi_k] = 
\frac{1}{2}
(\partial_\mu \Phi)^2 - \frac{m_\phi^2}{2}\Phi^2 -
\frac{\lambda}{4 !} \Phi^4 
+ \sum_{k=1}^{N_{\psi}}  
\bar{\psi}_{k} \left[i \not\!\partial - m_{\psi_k} -h_{k} \Phi
\right] \psi_{k}
\:.
\label{Lphipsi}
\end{eqnarray}
The $\Phi$ field is decomposed into a classical background
component $\varphi$ and a quantum fluctuation part as
$\Phi = \varphi + \phi$, with

\[
\langle \Phi \rangle = \varphi \;, \;\; \langle \phi \rangle =0 ,
\]

\noindent
and where the background component is
assumed to be homogeneous $\varphi \equiv \varphi(t)$.
Using the tadpole method\footnote{{}For earlier references on the tadpole
method applied to determining the equation of motion of a scalar field
in the real time formalism, see for instance 
Ref. \cite{ringwald}.}, where we impose
$\langle \phi \rangle = 0$ at all orders in perturbation
theory, it
leads to the condition that the sum of all tadpole terms
vanish.     
The effective equation of 
motion (EOM) for $\varphi(t)$ then becomes

\begin{eqnarray}
\ddot{\varphi}(t) + m_\phi^2 \varphi(t) + \frac{\lambda}{6} \varphi^3(t) 
+\frac{\lambda}{2} \varphi(t) \langle \phi^2 \rangle 
+\frac{\lambda}{6} \langle \phi^3 \rangle
+\sum_{k=1}^{N_{\psi}} h_{k} \langle \bar{\psi_k} \psi_k \rangle= 0 \;,
\label{eq1phi1}
\end{eqnarray}

\noindent 
where $\langle \phi^2 \rangle$,  $\langle \phi^3 \rangle$ 
and $\langle \bar{\psi}_k \psi_k \rangle$
are given in terms of the coincidence limit of the (causal) two-point
Green's functions $G^{++}_\phi (x,x')$ and $S^{++}_{\psi} (x,x')$.
These Green's functions are 
obtained from the $(1,1)$-component of the real time matrix of full
propagators which satisfy the appropriate Schwinger-Dyson equations
(see, {\it e.g.}, \cite{ian,BGR} for further details)

\begin{eqnarray}
\left[\Box + m_\phi^2 + \frac{\lambda}{2} \varphi^2 \right] 
G_\phi (x,x')
+ \int d^4 z \Sigma_\phi (x,z) G_\phi (z,x') = i \delta
(x,x')  
\label{Gphi}
\end{eqnarray}

\noindent
and

\begin{eqnarray}
\left[ 
i \not\!\partial - m_{\psi_k} -h_k \varphi
\right] S_{\psi_k} (x,x') 
+ \int d^4 z \Sigma_{\psi_k} (x,z) S_{\psi_k} (z,x') = i \delta
(x,x')\;.  
\label{Spsi}
\end{eqnarray}

\noindent
The momentum-space {}Fourier 
transform
of $G_\phi (x,x')$ (for the scalar field) can be expressed in the form

\begin{equation}
G_\phi(x,x') =i \int \frac{d^3 q}{(2 \pi)^3} 
e^{i {\bf q} . ({\bf x} - {\bf x}')}
\left(
\begin{array}{ll}
G^{++}_\phi({\bf q}, t- t') & \:\: G^{+-}_\phi({\bf q}, t-t') \\
G^{-+}_\phi({\bf q}, t- t') & \:\: G^{--}_\phi({\bf q}, t-t')
\end{array}
\right) \: ,
\label{Gmatrix}
\end{equation}

\noindent
where

\begin{eqnarray}
\lefteqn{G^{++}_\phi({\bf q} , t-t') = G^{>}_\phi({\bf q},t-t')
\theta(t-t') + G^{<}_\phi({\bf q},t-t') \theta(t'-t)} ,
\nonumber \\
& & G^{--}_\phi({\bf q} , t-t') = G^{>}_\phi({\bf q},t-t')
\theta(t'-t) + G^{<}_\phi({\bf q},t-t') \theta(t-t') ,
\nonumber \\
& & G^{+-}_\phi({\bf q} , t-t') = G^{<}_\phi({\bf q},t-t') ,
\nonumber \\
& & G^{-+}_\phi({\bf q},t-t') = G^{>}_\phi({\bf q},t-t')\;.
\label{G of k}
\end{eqnarray}

\noindent
In these expressions
the fully dressed (field independent) two-point functions, 
at finite temperature $T=1/\beta$,
under the approximation that the spectral function for the scalar field
has the standard Breit-Wigner form, are given by\footnote{We 
thank Ian Lawrie for pointing out the correct form of these equations.}

\begin{eqnarray} 
G^{>}_\phi({\bf q} ,t-t') &= & 
\frac{1}{2\omega_\phi}\left\{[1+n_\phi(\omega_\phi-i\Gamma_\phi)] 
e^{-i(\omega_\phi-i\Gamma_\phi)(t-t')} 
+ n_\phi(\omega_\phi+i\Gamma_\phi)e^{i(\omega_\phi+i\Gamma_\phi)(t-t')} 
\right\}\theta(t-t') 
\nonumber\\ 
&+&\frac{1}{2\omega_\phi}\left\{[1+n_\phi(\omega_\phi+i\Gamma_\phi)] 
e^{-i(\omega_\phi+i\Gamma_\phi)(t-t')} 
+ n_\phi(\omega_\phi-i\Gamma_\phi)e^{i(\omega_\phi-i\Gamma_\phi)(t-t')} 
\right\}\theta(t'-t)\;, 
\nonumber \\ 
G^{<}_\phi({\bf q} , t-t') & = & G^{>}_\phi({\bf q}, t'-t) \: , 
\label{G><} 
\end{eqnarray}

\noindent 
where $n_\phi$ is the Bose 
distribution function,  
$\omega_\phi\equiv \omega_\phi({\bf q})$ is the particle's  
dispersion relation and $\Gamma_\phi$ 
is the $\phi$ decay width, defined as usual in terms of the field  
self-energy by

\begin{equation} 
\Gamma_\phi (q) =   
\frac{{\rm Im} \Sigma_\phi ({\bf q},\omega_\phi)} 
{2 \omega_\phi}\;. 
\label{Gammaphi} 
\end{equation} 
 
{}For the fermion fields we have instead that

\begin{eqnarray} 
&& S^{++}_\psi({\bf q},t-t')=S^{>}_\psi({\bf q},t-t')\theta(t-t') 
+S^{<}_\psi({\bf q},t-t')\theta(t'-t) , \nonumber\\ 
&& S^{--}_\psi({\bf k},t-t')=S^{>}_\psi({\bf q},t-t')\theta(t'-t) 
+S^{<}_\psi({\bf q},t-t')\theta(t-t') , \nonumber\\ 
&& S^{+ -}_\psi({\bf q},t-t')= 
S^{<}_\psi({\bf q},t-t') , \nonumber\\ 
&& S^{- +}_\psi({\bf q},t-t')= 
S^{>}_\psi({\bf q},t-t')\;. 
\label{Gferm} 
\end{eqnarray} 
 
\noindent 
In this case $S^{>,<}_\psi$ are given by\footnote{{}Analogous expressions for 
the general nonequilibrium case are given in \cite{ian2}.}

\begin{eqnarray} 
&& S^{>}_\psi({\bf q},t-t')  
=-\frac{1}{2\omega_\psi}\left\{e^{-i(\omega_\psi-i\Gamma_\psi)(t-t')} 
\left[(\omega_\psi-i \Gamma_\psi) \gamma_0 - 
\not\!{\bf q}+m_{\psi,r} - 
i\frac{\Gamma_\psi(\omega_\psi - i \Gamma_\psi)}{m_{\psi,r}}\right] 
\right.\nonumber \\  
&& \left. \times \left[1-n_\psi(\omega_\psi-i\Gamma_\psi) 
\right]\right. \nonumber \\ 
&&\left. -e^{i(\omega_\psi+i\Gamma_\psi) (t-t')} 
\left[-(\omega_\psi+i\Gamma_\psi)\gamma_0-\not\!{{\bf q}}+m_{\psi,r}+ 
i\frac{\Gamma_\psi(\omega_\psi + i \Gamma_\psi)}{m_{\psi,r}}\right]  
n_\psi(\omega_\psi + i \Gamma_\psi)  \right\} \theta(t-t') 
\nonumber \\ 
&& -\frac{1}{2\omega_\psi}\left\{e^{-i(\omega_\psi+i\Gamma_\psi)(t-t')} 
\left[(\omega_\psi+i \Gamma_\psi) \gamma_0 - 
\not\!{\bf q}+m_{\psi,r} + 
i\frac{\Gamma_\psi(\omega_\psi + i \Gamma_\psi)}{m_{\psi,r}}\right] 
\right.\nonumber \\  
&& \left.\times  \left[1-n_\psi(\omega_\psi+i\Gamma_\psi) 
\right]\right. \nonumber \\ 
&&\left. -e^{i(\omega_\psi-i\Gamma_\psi) (t-t')} 
\left[-(\omega_\psi-i\Gamma_\psi)\gamma_0-\not\!{{\bf q}}+m_{\psi,r}- 
i\frac{\Gamma_\psi(\omega_\psi - i \Gamma_\psi)}{m_{\psi,r}}\right]  
n_\psi(\omega_\psi - i \Gamma_\psi)  \right\} \theta(t'-t) 
\;, 
\label{Spsi>} 
\end{eqnarray} 
 
\noindent 
and  
 
\begin{eqnarray} 
&& S^{<}_\psi({\bf q},t-t')  
=\frac{1}{2\omega_\psi}\left\{e^{i(\omega_\psi+i\Gamma_\psi)(t-t')} 
\left[-(\omega_\psi+i \Gamma_\psi) \gamma_0 - 
\not\!{\bf q}+m_{\psi,r} + 
i\frac{\Gamma_\psi(\omega_\psi + i \Gamma_\psi)}{m_{\psi,r}}\right] 
\right.\nonumber \\  
&& \left. \times \left[1-n_\psi(\omega_\psi+i\Gamma_\psi) 
\right]\right. \nonumber \\ 
&&\left. -e^{-i(\omega_\psi-i\Gamma_\psi) (t-t')} 
\left[(\omega_\psi-i\Gamma_\psi)\gamma_0-\not\!{{\bf q}}+m_{\psi,r}- 
i\frac{\Gamma_\psi(\omega_\psi - i \Gamma_\psi)}{m_{\psi,r}}\right]  
n_\psi(\omega_\psi - i \Gamma_\psi)  \right\} \theta(t-t') 
\nonumber \\ 
&& +\frac{1}{2\omega_\psi}\left\{e^{i(\omega_\psi-i\Gamma_\psi)(t-t')} 
\left[-(\omega_\psi-i \Gamma_\psi) \gamma_0 - 
\not\!{\bf q}+m_{\psi,r} - 
i\frac{\Gamma_\psi(\omega_\psi - i \Gamma_\psi)}{m_{\psi,r}}\right] 
\right.\nonumber \\  
&& \left. \times \left[1-n_\psi(\omega_\psi-i\Gamma_\psi) 
\right]\right. \nonumber \\ 
&&\left. -e^{-i(\omega_\psi+i\Gamma_\psi) (t-t')} 
\left[(\omega_\psi+i\Gamma_\psi)\gamma_0-\not\!{{\bf q}}+m_{\psi,r}+ 
i\frac{\Gamma_\psi(\omega_\psi + i \Gamma_\psi)}{m_{\psi,r}}\right]  
n_\psi(\omega_\psi + i \Gamma_\psi)  \right\} \theta(t'-t) 
\;, 
\label{Spsi<} 
\end{eqnarray}

\noindent 
where $n_\psi = 1/(e^{\beta \omega_\psi}+1)$, $\omega_\psi= 
\sqrt{{\bf q}^2+m_{\psi,r}^2}$, with $m_{\psi,r}=m_\psi +  
{\rm Re}\Sigma_\psi$ and  
 
\begin{equation} 
\Gamma_\psi(q) = {\rm Im} \Sigma_\psi(\omega_\psi,{\bf q})  
\frac{m_\psi}{\omega_\psi}. 
\label{Gammapsi} 
\end{equation}

\noindent 
In this paper we will be interested in studying 
dissipation only of the background scalar field $\varphi$ at $T=0$. 
Therefore we will restrict ourselves just to the zero temperature  
expressions of the above equations. 
  
It will be assumed the couplings 
$\lambda,h_k \ll 1$, so that   
perturbation theory 
can be consistently formulated with subleading terms neglected. 
Then by perturbatively expanding the field averages in Eq. (\ref{eq1phi1}),  
up to two-loop order in the scalar 
loops (higher order loop terms involving fermions also 
are higher order in the perturbation expansion,  
as we will discuss later), the effective EOM 
for $\varphi(t)$ is 
 
\begin{eqnarray} 
\lefteqn{\ddot{\varphi}(t) + \left( m_\phi^2 +\frac{\lambda}{2} 
\langle\phi^2\rangle_0 \right) \varphi (t) +  
\frac{\lambda}{6}  
\varphi^3 (t) } \nonumber \\ 
& & + \lambda \varphi (t) \int_{-\infty}^t dt' \: 
\frac{\lambda}{2} \varphi^2 (t')  \int \frac{d^3 {\bf q}}{(2 \pi)^3}  
{\rm Im} \left[ G_\phi^{++} ({\bf q},t-t') \right]^2 \nonumber \\ 
& & + \frac{\lambda^2}{3} \int_{-\infty}^{t} d t'  
\varphi (t') \int \frac{d^3 {\bf q}_1}{(2 \pi)^3}  
\frac{d^3 {\bf q}_2}{(2 \pi)^3} 
{\rm Im} \left[ G_\phi^{++} ({\bf q}_1,t-t') G_\phi^{++} ({\bf q}_2,t-t')   
G_\phi^{++} ({\bf q}_1+{\bf q}_2,t-t')\right] \nonumber \\ 
& & - 4 \sum_{k=1}^{N_\psi} h_{k,\phi}  
\int_{-\infty}^{t} dt'\; h_{k} \varphi (t') \int \frac{d^3 q}{(2 \pi)^3} 
{\rm Im} \left[ S_{\alpha \beta}^{++} ({\bf q},t-t')  
S^{++\, \beta \alpha} ({\bf q},t'-t) \right] = 0\;, 
\label{eq1phi2} 
\end{eqnarray}

\noindent  
with $\langle \ldots\rangle_0$ meaning the  
vacuum expectation value at zero background field 
$\varphi=0$. Using the explicit expressions for the  
Green's functions in Eq. (\ref{eq1phi2})  
and at $T=0$, this 
EOM becomes 
 
\begin{eqnarray} 
&& \ddot{\varphi}(t) + \left( m_\phi^2 +\frac{\lambda}{2} 
\langle\phi^2\rangle_0 \right) \varphi (t) +  
\frac{\lambda}{6}  
\varphi^3 (t) 
- \lambda^2 \varphi (t) \int_{-\infty}^t dt'\; 
\varphi^2 (t') \int \frac{d^3 q}{(2 \pi)^3} 
\frac{\sin(2 \omega_\phi |t-t'|)}{8 \omega_\phi^2} e^{-2 \Gamma_\phi(q) 
|t-t'|}\nonumber \\ 
& & - \frac{\lambda^2}{3} \int_{-\infty}^{t} d t'  
\varphi (t') \int \frac{d^3 {\bf q}_1}{(2 \pi)^3}  
\frac{d^3 {\bf q}_2}{(2 \pi)^3} 
\frac{\sin[(\omega_\phi ({\bf q}_1) + \omega_\phi ({\bf q}_2) 
+\omega_\phi ({\bf q}_1+{\bf q}_2) )|t-t'|]}{8 \omega_\phi ({\bf q}_1)  
\omega_\phi ({\bf q}_2) \omega_\phi ({\bf q}_1+{\bf q}_2)} \nonumber \\ 
&& \times 
e^{-[\Gamma_\phi ({\bf q}_1) + \Gamma_\phi ({\bf q}_2) + 
\Gamma_\phi ({\bf q}_1+{\bf q}_2) ] |t-t'|} \nonumber \\ 
& & - 2 \sum_{k=1}^{N_\psi} h_k^2 \;  
\int_{-\infty}^t  dt' \;\varphi (t') \int \frac{d^3 q}{(2 \pi)^3} 
\frac{e^{-2 \Gamma_{\psi_k}(q) |t-t'|}}{\omega_{\psi_k}^2 m_{\psi,r}^2} 
\left\{ 2 \Gamma_{\psi_k}^3 \omega_{\psi_k}    
\cos(2 \omega_{\psi_k} |t-t'|)  \right. \nonumber \\ 
&& \left. + \left[ 2 m_{\psi,r}^2 (\omega_{\psi_k}^2 - m_{\psi_k}^2)+ 
\Gamma_{\psi_k}^2 (\omega_{\psi_k}^2+ m_{\psi,r}^2-\Gamma_{\psi_k}) 
\right] \sin(2 \omega_{\psi_k} |t-t'|) \right\} 
= 0\;. 
\label{eq1phi3} 
\end{eqnarray}

The temporally nonlocal terms of the types appearing in  
the above equation 
have been shown in \cite{boya,GR,BGR,hu,hu4,greiner,RF} to 
lead to dissipative dynamics in the EOMs. This can be made more explicit 
by an appropriate 
integration by parts in the time integrals in Eq. (\ref{eq1phi3}) 
(see \cite{boya,RF}) to obtain the result (at $T=0$) 
 
\begin{eqnarray} 
&& \ddot{\varphi}(t) + \bar{m}_{\phi}^2 \; \varphi (t) +  
\frac{\bar{\lambda}}{6}  
\varphi^3 (t) \nonumber \\ 
&&+ \lambda^2 \varphi (t) \int_{-\infty}^t dt'\; 
\varphi (t') \dot{\varphi} (t') \int \frac{d^3 q}{(2 \pi)^3} 
\frac{ \left[\omega_\phi \cos(2 \omega_\phi |t-t'|) + \Gamma_\phi  
\sin(2 \omega_\phi 
|t-t'|) \right] }{ 8 \omega_\phi^2 \left( \Gamma_\phi^2 + \omega_\phi^2 
\right) } e^{-2 \Gamma_\phi(q) |t-t'|} \nonumber \\ 
&& + \frac{\lambda^2}{3} \int_{-\infty}^{t} d t' \; 
\dot{\varphi} (t') \int \frac{d^3 {\bf q}_1}{(2 \pi)^3}  
\frac{d^3 {\bf q}_2}{(2 \pi)^3} 
\left\{ \left[\omega_\phi ({\bf q}_1) + \omega_\phi ({\bf q}_2) + 
\omega_\phi ({\bf q}_1+{\bf q}_2)\right] \right. \nonumber \\ 
&& \times \left. \cos[(\omega_\phi ({\bf q}_1) +  
\omega_\phi ({\bf q}_2) 
+\omega_\phi ({\bf q}_1+{\bf q}_2) )|t-t'|]  \right.  
\nonumber \\  
&& \left. + \left[\Gamma_\phi ({\bf q}_1) + \Gamma_\phi ({\bf q}_2) + 
\Gamma_\phi ({\bf q}_1+{\bf q}_2)\right] \sin[(\omega_\phi ({\bf q}_1) +  
\omega_\phi ({\bf q}_2) 
+\omega_\phi ({\bf q}_1+{\bf q}_2) )|t-t'|]\right\} \nonumber \\ 
&& \times \frac{ 
e^{-[\Gamma_\phi ({\bf q}_1) + \Gamma_\phi ({\bf q}_2) + 
\Gamma_\phi ({\bf q}_1+{\bf q}_2) ] |t-t'|}}{8 \omega_\phi ({\bf q}_1)  
\omega_\phi ({\bf q}_2) \omega_\phi ({\bf q}_1+{\bf q}_2)}  
\nonumber \\ 
&& \times \frac{1}{  \left[\Gamma_\phi ({\bf q}_1) +  
\Gamma_\phi ({\bf q}_2) + 
\Gamma_\phi ({\bf q}_1+{\bf q}_2) \right]^2 + \left[\omega_\phi ({\bf q}_1) + 
\omega_\phi ({\bf q}_2) +\omega_\phi ({\bf q}_1+{\bf q}_2)\right]^2  } 
\nonumber \\ 
& & + \sum_{k=1}^{N_\psi} h_k^2 \;  
\int_{-\infty}^t  dt' \; \dot{\varphi} (t')  
\int \frac{d^3 q}{(2 \pi)^3} 
\frac{e^{-2 \Gamma_{\psi_k} (q) |t-t'|}}{ \omega_{\psi_k}^2 m_{\psi,r}^2  
\left( \Gamma_{\psi_k}^2 + \omega_{\psi_k}^2 \right) } 
\nonumber \\   
&& \times \left\{ \left[ \Gamma_{\psi_k}^4 \omega_{\psi_k} +  
2 \omega_{\psi_k}^3 
m_{\psi,r}^2 - 2 m_{\psi,r}^4 \omega_{\psi_k} + \Gamma_{\psi_k}^2 
\omega_{\psi_k} (\omega_{\psi_k}^2 + m_{\psi,r}^2) \right] 
\cos(2 \omega_{\psi_k} |t-t'|) \right. \nonumber \\  
&& + \left. \left[ \Gamma_{\psi_k}(2 m_{\psi_k}^2 - \Gamma_{\psi_k}^2)  
(\omega_{\psi_k}^2 - m_{\psi,r}^2) -\Gamma_{\psi_k}^5 \right] 
\sin(2 \omega_{\psi_k}|t-t'|) \right\} 
= 0\;, 
\label{eq1phi4} 
\end{eqnarray}

\noindent 
where $\bar{m}_{\phi}$ and $\bar{\lambda}$ are the effective  
mass and coupling constant given respectively by 
 
\begin{eqnarray} 
\bar{m}_{\phi}^2 &=& m_\phi^2+ \lambda \int\frac{d^3 q}{(2 \pi)^3}  
\frac{1}{4 \omega_\phi}  +   \sum_{k=1}^{N_\psi} h_k^2  
\int\frac{d^3 q}{(2 \pi)^3} \left[2  \frac{m_{\psi,r}^2-\omega_{\psi_k}^2} 
{\omega_{\psi_k}^3} +  
{\cal O}\left(\frac{\Gamma_{\psi_k}^2}{\omega_{\psi_k}^2} \right) \right] 
\nonumber \\ 
&-& \frac{\lambda^2}{3} \int \frac{d^3 {\bf q}_1}{(2 \pi)^3}  
\frac{d^3 {\bf q}_2}{(2 \pi)^3} \left\{ \frac{1}{8 \omega_\phi ({\bf q}_1)  
\omega_\phi ({\bf q}_2) \omega_\phi ({\bf q}_1+{\bf q}_2) 
[\omega_\phi ({\bf q}_1) + 
\omega_\phi ({\bf q}_2) +\omega_\phi ({\bf q}_1+{\bf q}_2)]} \right. 
\nonumber \\ 
&+& \left. 
{\cal O}\left(\frac{\Gamma_{\phi}^2}{\omega_{\phi}^2} \right) \right\} 
\;, 
\label{massr} 
\end{eqnarray} 
 
\noindent 
and 
 
\begin{equation} 
\bar{\lambda} = \lambda - \lambda^2 \int\frac{d^3 q}{(2 \pi)^3}  
\left[ \frac{1}{16 \omega_\phi^3} +  
{\cal O}\left(\frac{\Gamma_{\phi}^2}{\omega_{\phi}^2} \right) \right]\;. 
\label{lambdar} 
\end{equation} 
 
\noindent 
Both $\bar{m}_{\phi}$ and $\bar{\lambda}$ naively 
appear divergent in Eqs. (\ref{massr}) and (\ref{lambdar}) 
respectively, due to the perturbative correction terms. 
However, they are rendered finite 
by the usual introduction of counterterms to renormalize 
the mass and coupling constant in the original Lagrangian 
\cite{GR}.  In this procedure, the bare (infinite) mass and scalar 
self-coupling in the Lagrangian are $m=m_{\phi,r}+\delta m_\phi$ and 
$\lambda=\lambda_r+\delta \lambda$, where the counterterms  
$\delta m_\phi$ and $\delta \lambda$ cancel the divergent contributions 
in Eqs. (\ref{massr}) and (\ref{lambdar}) in the usual  
way. In what follows, we then can just interpret the masses 
and couplings appearing in our equations as the renormalized ones 
and we will omit any additional subscript  
and overbars for simplicity, so 
hereafter  ${\bar m} \rightarrow m$ and 
${\bar \lambda} \rightarrow \lambda$. 
 
\subsection{The EOM in the adiabatic-Markovian approximation} 
\label{sect2A} 
 
The effective equation of motion 
Eq. (\ref{eq1phi4}) is the main result of this section. 
As should appear evident, this equation is difficult 
to solve either analytically or numerically, 
since memory of the past history of $\varphi$ is required at each 
stage of evolution.  {}Falling short of a detailed analysis of this 
equation in this paper, we would like some rough estimates 
of the dissipative effects described by the equation. 
In particular, as a first approximation to Eq. ({\ref{eq1phi4}), 
we will explore the derivative expansion at leading order 
for the temporally nonlocal terms.  This amounts to substituting 
$t' \rightarrow t$ in the arguments of the fields entering in 
the time integrals.  This is equivalent to a Markovian 
approximation for the dissipative kernels in Eq. (\ref{eq1phi4}).  
A solution  regime for $\varphi$ where such 
an approximation might be valid is the adiabatic regime, where 
the motion of $\varphi$ is slow.  In this  
and the next subsections, we will 
establish a self-consistent solution regime composed of both 
the adiabatic and Markovian approximations. 
To proceed, first the Markovian approximation will 
be implemented and then self-consistent solution 
regimes will be identified.  Thus upon implementing   
the Markovian approximation, we then can easily perform the 
$t'$ integrals and obtain 
 
\begin{eqnarray} 
&& \ddot{\varphi}(t) + m_{\phi}^2 \; \varphi (t) +  
\frac{\lambda}{6}  
\varphi^3 (t) 
+ \lambda^2 \varphi^2 (t)  
\dot{\varphi} (t) \int \frac{d^3 q}{(2 \pi)^3} 
\frac{\Gamma_\phi}{8 \omega_\phi  
(\omega_\phi^2+\Gamma_\phi^2)^2} \nonumber \\  
&& + \frac{\lambda^2}{3} \dot{\varphi}(t) \int \frac{d^3 {\bf q}_1}{(2 \pi)^3}  
\frac{d^3 {\bf q}_2}{(2 \pi)^3}   
\frac{\left[\Gamma_\phi ({\bf q}_1) +  
\Gamma_\phi ({\bf q}_2) + 
\Gamma_\phi ({\bf q}_1+{\bf q}_2) \right] \left[ 
\omega_\phi ({\bf q}_1) + 
\omega_\phi ({\bf q}_2) +\omega_\phi ({\bf q}_1+{\bf q}_2)\right]} 
{4 \omega_\phi ({\bf q}_1)  
\omega_\phi ({\bf q}_2) \omega_\phi ({\bf q}_1+{\bf q}_2)} 
\nonumber \\ 
&& \times \frac{1} 
{\left\{ \left[\Gamma_\phi ({\bf q}_1) +  
\Gamma_\phi ({\bf q}_2) + 
\Gamma_\phi ({\bf q}_1+{\bf q}_2) \right]^2 + \left[\omega_\phi ({\bf q}_1) + 
\omega_\phi ({\bf q}_2) +\omega_\phi ({\bf q}_1+{\bf q}_2)\right]^2 
\right\}^2} \nonumber \\ 
&& + 
\sum_{k=1}^{N_\psi} h_k^2 \dot{\varphi} (t) \;  
\int \frac{d^3 q}{(2 \pi)^3}  
\frac{\left(\Gamma_{\psi_k}^2 + 
2 \omega_{\psi_k}^2-2 m_{\psi_k}^2\right)}{ 
\omega_{\psi_k} (\Gamma_{\psi_k}^2+\omega_{\psi_k}^2)^2} \Gamma_{\psi_k} 
=0 \;. 
\label{eq1phi5} 
\end{eqnarray}

An additional simplification can be achieved by demanding that 
the (effective) field masses satisfy the inequality   
$m_\phi > 2 m_{\psi_k}$. In this case, at zero temperature, the only  
contribution to the decay widths in Eqs. (\ref{Gammaphi}) and 
(\ref{Gammapsi}) come from the imaginary part of the one loop 
contribution to the self-energy for $\Phi$.  This is given by the 
internal fermionic propagators diagram, which then represents 
the decay rate for the kinematically allowed process  
$\Phi \to \psi_k + \bar{\psi}_k$. We therefore have that  
$\Gamma_{\psi_k} =0$ and $\Gamma_\phi$ is given by (see the Appendix)

\begin{equation} 
\Gamma_\phi (q) =  
\sum_{k=1}^{N_\psi}  
\frac{h_{k}^2}{8 \pi \omega_{\phi}({\bf q})} m_{\phi}^2 
\left(1 - \frac{4 m_{\psi_k}^2}{m_{\phi}^2} \right)^{\frac{3}{2}}\;. 
\label{Gamma1} 
\end{equation} 
 
We must point out that even though in this case the contribution 
to the dissipative term in Eq. (\ref{eq1phi5}) coming from the  
fermionic loop vanishes, i.e. $\Gamma_{\psi_k}=0$, this is a  
consequence of assuming the Markovian approximation for the 
dissipative kernels. In the general case, the last term 
in the non-local EOM, Eq. (\ref{eq1phi4}), with $\Gamma_{\psi_k}=0$, 
still can be interpreted as  a dissipative contribution \cite{boya,hu,hu4} 
to the EOM.  
Therefore, here the Markovian approximation somewhat under-estimates 
the whole dissipative nature of the non-local kernels in 
(\ref{eq1phi4}). We expect this not to invalidate our 
main objective here, which is to determine the viability of  
adiabatic dissipative behavior in the background field evolution and  
the intrinsic dissipative nature of the field dynamics. 
 
As an aside, it is interesting to discuss briefly the interpretation 
of dissipation coming from the second non-local dissipative term 
in Eq. (\ref{eq1phi4}) when $\Gamma_{\psi}=0$. 
Though for this case,  
particle decay does not contribute to dissipation, there still is  
dissipation due to ``off-shell'' excitation of 
virtual states which leads to decoherence and power law decay 
of the background field amplitude \cite{boya}, in the  
absence of the scalar field self-interactions. This contrasts with 
the first dissipative term in Eq. (\ref{eq1phi4}), 
in which dissipation truly is 
coming from the real scattering by the quasi-particles in the medium. 
{}Furthermore, dissipative kernels of the sort 
treated here typically have some type of long time 
tail, which retains memory from the past. 
However, as will be discussed later,  
a non-vanishing decay width, as present in this case, 
helps to suppress the long time tail of the kernel. 
 
Studies of terms similar to the second nonlocal  
term in Eq. (\ref{eq1phi4}) have been done in 
\cite{hu2,baacke,boya2}.  These works studied the 
case $N_{\psi}=1$ and $\Gamma_{\psi} = 0$, and examined 
the linearized form of the effective EOM once the fermion 
fields were integrated out. 
An important 
issue raised in \cite{baacke,boya2} was in regards 
the singularities in the EOM at the initial time, which in our  
case refers to the behavior of the dissipative kernel in Eq. (\ref{eq1phi4}) 
arising from the fermionic loop, when computed at $t=t'$.  Observe 
that this term 
appears divergent in the ultraviolet region, but 
this apparent problem easily can be solved  
in our approach. To see that our final equation in this case is  
identical to the final result obtained by the authors in \cite{baacke,boya2}, 
take for example the second nonlocal term in Eq. (\ref{eq1phi4}), due 
to the fermionic loop, and integrate the time  
integral twice by parts to give 
 
\begin{eqnarray} 
& & \sum_{k=1}^{N_\psi} h_k^2 \;  
\int_{-\infty}^t  dt' \; \dot{\varphi} (t')  
\int \frac{d^3 q}{(2 \pi)^3} 
\frac{e^{-2 \Gamma_{\psi_k} (q) |t-t'|}}{ \omega_{\psi_k}^2 m_{\psi_k}^2  
\left( \Gamma_{\psi_k}^2 + \omega_{\psi_k}^2 \right) } 
\nonumber \\   
&& \times \left\{ \left[ \Gamma_{\psi_k}^4 \omega_{\psi_k} +  
2 \omega_{\psi_k}^3 
m_{\psi_k}^2 - 2 m_{\psi_k}^4 \omega_{\psi_k} + \Gamma_{\psi_k}^2 
\omega_{\psi_k} (\omega_{\psi_k}^2 + m_{\psi_k}^2) \right] 
\cos(2 \omega_{\psi_k} |t-t'|) \right. \nonumber \\  
&& + \left. \left[ \Gamma_{\psi_k}(2 m_{\psi_k}^2 - \Gamma_{\psi_k}^2)  
(\omega_{\psi_k}^2 - m_{\psi_k}^2) -\Gamma_{\psi_k}^5 \right] 
\sin(2 \omega_{\psi_k}|t-t'|) \right\} \nonumber \\ 
&&= \sum_{k=1}^{N_\psi} h_k^2 \; \dot{\varphi} (t)  
\int \frac{d^3 q}{(2 \pi)^3} \frac{\left(\Gamma_{\psi_k}^2 + 
2 \omega_{\psi_k}^2-2 m_{\psi_k}^2\right)}{ 
\omega_{\psi_k} (\Gamma_{\psi_k}^2+\omega_{\psi_k}^2)^2} \Gamma_{\psi_k} 
\nonumber \\ 
&& + \sum_{k=1}^{N_\psi} h_k^2 \; \ddot{\varphi} (t)  
\int \frac{d^3 q}{(2 \pi)^3} \left[ \frac{\omega_{\psi_k}^2-m_{\psi_k}^2}{ 
2 \omega_{\psi_k}^5}  + {\cal O}\left( \frac{\Gamma_{\psi_k}^2} 
{\omega_{\psi_k}^2}\right) \right]  \nonumber \\ 
&& - \sum_{k=1}^{N_\psi} h_k^2 \; 
\int_{-\infty}^t  dt' \; \varphi \!\!\! \dot{}\; \dot{}\; \dot{} \,(t')  
\int \frac{d^3 q}{(2 \pi)^3} 
e^{-2 \Gamma_{\psi_k} (q) |t-t'|} 
\left\{ \frac{\omega_{\psi_k}^2-m_{\psi_k}^2}{ 
2 \omega_{\psi_k}^5} \left[ \cos(2 \omega_{\psi_k} |t-t'|) \right. \right. 
\nonumber \\ 
&& \left. \left. + 
3 \frac{\Gamma_{\psi_k}}{\omega_{\psi_k}} \sin(2 \omega_{\psi_k} |t-t'|) 
+ {\cal O}\left( \frac{\Gamma_{\psi_k}^2} 
{\omega_{\psi_k}^2}\right) \right]  \right\}\;. 
\label{third} 
\end{eqnarray} 
 
\noindent 
The first term on the rhs of (\ref{third}) is just the  
second dissipative term appearing in Eq. (\ref{eq1phi5}). 
By taking $\Gamma_{\psi_k}=0$ this term vanishes. The second term is  
a wave-function renormalization term, which can be absorbed  
in a wave function counterterm ($\delta Z$) in the Lagrangian, 
by rewriting the kinetic term as $(\partial_\mu \Phi)^2 \to 
(1+\delta Z) (\partial_\mu \Phi)^2$. The last term on the rhs of  
Eq. (\ref{third}), for $\Gamma_{\psi_k}=0$ and $N_\psi =1$,  
reproduces exactly the  
result obtained in \cite{baacke,boya2} for the EOM integrated over 
the fermion fields. Thus, we see that there is no ambiguities or divergences 
here associated with the kernels evaluated at equal times. 
Also, the non-local two-loop scalar term in Eq. (\ref{eq1phi4}) 
can be examined in a similar way and from such an analysis 
its contribution to the wave-function renormalization can be extracted. 
    
Returning from this digression to the EOM Eq. (\ref{eq1phi5}), and using  
Eq. (\ref{Gamma1}) for $\Gamma_\phi$ and $\Gamma_\psi=0$, 
the EOM becomes

\begin{eqnarray} 
&& \ddot{\varphi}(t) + m_{\phi}^2 \; \varphi (t) +  
\frac{\lambda}{6} \varphi^3 (t) + \eta (\varphi)  
\dot{\varphi} (t) =0\;. 
\label{final1} 
\end{eqnarray} 
 
\noindent 
Here $\eta(\varphi)$ is the dissipative coefficients, which 
after doing the momentum integral in the ``one-loop'' dissipation term 
and using the symmetry of the ``two-loop'' dissipation term under  
change of momentum integration variables, becomes 
 
\begin{eqnarray} 
\eta (\varphi) &=&\varphi^2 (t) \frac{\lambda^2 \alpha_{\phi,\psi}^2 } 
{128 \pi \; \sqrt{m_{\phi}^4 +  
\alpha_{\phi,\psi}^4}\; \left(2 \sqrt{m_{\phi}^4 + \alpha_{\phi,\psi}^4} + 
2m_{\phi}^2 \right)^{\frac{1}{2}}}\nonumber \\ 
&+&  \frac{\lambda^2 \alpha_{\phi,\psi}^2}{4}  
\int \frac{d^3 {\bf q}_1}{(2 \pi)^3}  
\frac{d^3 {\bf q}_2}{(2 \pi)^3}  
\left\{ 
\frac{1}{\omega_\phi ({\bf q}_1)^2  
\omega_\phi ({\bf q}_2) \omega_\phi ({\bf q}_1+{\bf q}_2) 
\left[\omega_\phi ({\bf q}_1) + 
\omega_\phi ({\bf q}_2) +\omega_\phi ({\bf q}_1+{\bf q}_2)\right]^3}  
\right. \nonumber \\ 
&+& \left. {\cal O}\left(\frac{\Gamma_{\phi}^2}{\omega_{\phi}^2} \right)  
\right\}\;, 
\label{eta1} 
\end{eqnarray} 
 
\noindent 
with 
 
\begin{equation} 
\alpha_{\phi,\psi}^2 = \sum_{k=1}^{N_\psi} \frac{h_{k}^2}{8 \pi} m_{\phi}^2 
\left(1-\frac{4 m_{\psi_k}^2}{m_{\phi}^2} \right)^{\frac{3}{2}} \;. 
\label{alpha} 
\end{equation}

{}Finally, we also can easily work out an  
equivalent model of $\Phi$ coupled to bath fields that 
rather than fermionic fields are scalar fields, 
$\chi_j$, $j=1 \ldots N_\chi$, with a  
trilinear coupling as 
$\sum_{j=1}^{N_{\chi}}\frac{g_{j}^2}{2} \Phi \chi_{j}^2$. 
Once again, by choosing $m_\phi > 2 m_{\chi_j}$ we have  
$\Gamma_{\chi_j}=0$ but a nonvanishing  $\Phi$ decay width, 
$\Gamma_\phi$, which also has been evaluated in the Appendix. 
Going through the same steps as used to obtain Eq. (\ref{final1}), 
we obtain an analogous expression with a similar dissipative coefficient 
$\eta(\varphi)$, except with $\alpha_{\phi,\psi} \to \alpha_{\phi,\chi}$,  
where 
 
\begin{equation} 
\alpha_{\phi,\chi}^2 = \sum_{j=1}^{N_\chi} \frac{g^4}{16 \pi}  
\left(1-\frac{4 m_{\chi_j}^2}{m_{\phi}^2} \right)^{\frac{1}{2}} \;. 
\label{alphachi} 
\end{equation} 
 
\subsection{Examination of the dissipative kernels} 
\label{sect2B} 
 
Up to this point, the naive 
implementation of the Markovian approximation 
has been examined and has led  
from Eq. (\ref{eq1phi4}) to Eq. (\ref{final1}).  
The applicability of this approximation is now considered.    
Returning to the non-local EOM Eq. (\ref{eq1phi4}), 
we can express for example the term generating the one-loop dissipation  
contribution in Eq. (\ref{final1}) 
in the general form\footnote{Although here we analyze the 
one-loop dissipative kernel in detail,  
similar conclusions also can be shown to apply for the more 
complicate two-loop dissipative kernel in Eq. (\ref{eq1phi4}).} 
 
\[ 
\int_{-\infty}^{t} dt' \:  \varphi(t') \dot{\varphi}(t') K(t,t')\;, 
\] 
 
\noindent 
where the dissipative kernel is given by 
 
\begin{equation} 
K(t,t') = \lambda^2 \int \frac{d^3 q}{(2 \pi)^3} 
\frac{ \left[\omega_\phi \cos(2 \omega_\phi |t-t'|) + \Gamma_\phi  
\sin(2 \omega_\phi 
|t-t'|) \right] }{ 8 \omega_\phi^2 \left( \Gamma_\phi^2 + \omega_\phi^2 
\right) } e^{-2 \Gamma_\phi(q) |t-t'|}\;. 
\label{kernel} 
\end{equation} 
 
\noindent 
In {}Fig. 1 $K(t,t'=0)/\lambda^2$ is plotted for four 
cases, $\Gamma_{\phi}(0)/m_{\phi} = 0.1,0.5,1.0,5.0$. 
There is a degeneracy amongst the parameters $N_{\psi}$, 
$h$ and $m_{\psi}$ from which to choose for a given 
value of $\Gamma_{\phi}(0)$.  The plots of the kernel 
are given over four different time interval, 
in order to see the different aspects of its behavior. 
 
As can be seen, the kernel has a 
pronounced, narrow peak around $t=t'$ 
followed by a power-law  
decaying oscillatory behavior, with greater decaying 
as the number of fields coupled to  
$\Phi$ increases.  Since from Eq. (\ref{Gamma1}) $\Gamma_{\phi}  
\sim 1/|{\bf q}|$ for 
$|{\bf q}| \gg m_{\phi}$, it can be seen from Eq. (\ref{kernel}) 
that once $|{\bf q}| \gtrsim 2 m_{\phi} \Gamma(0)|t-t'|$, damping 
from the exponential term relinquishes. 
Thus for $|t-t'| \gtrsim 1/m_{\phi}$ 
the behavior of the kernel is $K(t,t') \sim - \lambda^2  
{\rm Ci} (4 m_\phi \Gamma_\phi (0) |t-t'|^2) \sim - \lambda^2 
\sin  (4 m_\phi \Gamma_\phi (0) |t-t'|^2)/  
(4 m_\phi \Gamma_\phi (0) |t-t'|^2)$. 
Note this long time behavior of the $T=0$ contribution 
of the kernel differs from its high temperature component 
which was studied in \cite{GR,RF}.  At high temperature, the  
decay widths at large $|{\bf q}|$  
behave as $\sim T^2/|{\bf q}|$, but 
there are also factors of the number density which 
become Boltzmann suppressed. As such, the high temperature 
limit of the kernel becomes highly  exponentially suppressed at  
large times. 
 
Power-law decay of the kernel at $T=0$ implies that memory 
of the scalar field is retained in Eq. (\ref{eq1phi4}) in determining 
its future evolution.  As such, the derivative expansion clearly 
is not generally valid. However, for sufficiently slow 
motion of the scalar field, the derivative expansion 
still may be valid for some duration of time.  In particular, suppose 
${\dot \varphi}/{\varphi} \approx \gamma$ in a given solution, 
where $\gamma$ is the approximate magnitude of this ratio 
over some interval of time.  Then, 
self-consistency of this solution with respect to 
the derivative expansion  holds for a time 
interval $|t-t'| \lesssim 1/\gamma$. 
Thus, as mentioned at the beginning of 
this subsection, the slower is the motion 
of $\varphi$, the longer the above approximation is 
valid.  Such slowly varying solutions are useful to investigate 
due to their simplicity. They also may have 
practical use for example for warm inflation, 
where one seeks solutions where the motion of $\varphi$ is slow. 

Although the kernel Eq. (\ref{kernel}) 
does retain past memory, it  
is worth noting that on general grounds for reasonable 
$\varphi(t)$ solution regimes, 
the memory retention only is up to some finite time 
in the past.  The observation here is 
that the oscillation rate of the kernel  
increases with time due to the quadratic dependence on time 
$K(t,t') \sim \sin (a |t-t'|^2)$, 
where $a$ is a constant.  As such, for reasonable motions of 
$\varphi$, beyond some time interval into the past 
$|t-t'| > \Delta t_0$, 
the characteristic oscillation frequency of the kernel will 
exceed that of $\varphi$.  Thus at all times past this point, 
the contributions from $\varphi(t)$ primarily cancel. 
As a practical point, despite this property of the kernel to
filter through increasingly high frequency components as the time
passes, there are limitations to the types of
$\varphi(t)$ motions which it can describe. In particular,
this feature of the kernel most efficiently is able to cut-off 
the long time
memory if the field configuration only has slow frequency components.
As such, ideally the applicability of this kernel is
near equilibrium conditions.  Note, this also is
the regime assumed in previous works \cite{GR,BGR} for the finite
temperature case.

An additional consistency condition related to the adiabatic approximation 
at finite $T$, considered in \cite{BGR,BGR2}, is that the microscopic 
dynamics should be much faster than all macroscopic 
motions. {}For the finite-$T$ case, this was a necessary requirement 
since it guaranteed the system thermalized fast enough to 
adjust to any changes in the macroscopic state. The need for rapid 
thermalization was necessary for self-consistency of the solutions. 
In contrast, the $T=0$ dynamics treated in this paper 
does not require a specific statistical state in which the 
scalar field evolution occurs. Thus, consistency requirements, 
if any, with respect to the rates of microscopic versus macroscopic 
dynamics are less well defined for the $T=0$ case. 
To properly address this question, a complete nonequilibrium 
analysis is necessary, which is beyond the scope of this paper. 
However, for the time being, to be conservative, a similar 
consistency condition to the finite-$T$ case adopted in 
\cite{BGR} will be imposed, which requires the rate of microscopic 
physics to be faster than all macroscopic motions. 
Since the only scales characterizing the microscopic physics 
are the decay rates, this requirement implies 
 
\begin{equation} 
\left| \frac{\varphi}{\dot{\varphi}} \right| \gg \Gamma^{-1} , 
\label{kinetic} 
\end{equation} 
 
\noindent 
which is analogous to the condition in \cite{BGR}, except above the 
zero temperature decay rates are used.  In Sec. \ref{sect4} some estimates 
will be given of dissipative dynamics based on the combined 
adiabatic-Markovian approximation of this and the previous subsections.

\section{Indirect decay Models}  
\label{sect3} 
 
This section will consider models in which the particles ultimately 
created from dissipation of the scalar field are coupled indirectly to 
the scalar field through an intermediate field.  Such a case has 
much more variety in the types of decay sequences, 
as compared to the direct decay models of Sec. \ref{sect2}. 
The basic model to be examined consists of the system, a 
scalar field $\Phi$, along with scalar fields  $\chi_j$, $j=1 \ldots  
N_\chi$ and fermion fields $\psi_k$, $k=1 \ldots N_\psi$. 
The Lagrangian density is given by 
 
\begin{eqnarray} 
{\cal L} [ \Phi, \chi_j, \bar{\psi}_k, \psi_k] &=&  
\frac{1}{2} 
(\partial_\mu \Phi)^2 - \frac{m_\phi^2}{2}\Phi^2 - 
\frac{\lambda}{4 !} \Phi^4  
+ \sum_{j=1}^{N_{\chi}} \left\{ 
\frac{1}{2} (\partial_\mu \chi_{j})^2 - \frac{m_{\chi_j}^2}{2}\chi_j^2 
- \frac{f_{j}}{4!} \chi_{j}^4 - \frac{g_{j}^2}{2} 
\Phi^2 \chi_{j}^2  
\right\} 
\nonumber \\ 
&+& \sum_{k=1}^{N_{\psi}}   
\bar{\psi}_{k} \left[i \not\!\partial - m_{\psi_k} -h_{k,\phi} \Phi 
- \sum_{j=1}^{N_\chi} h_{kj,\chi} \chi_j \right] \psi_{k} 
\: , 
\label{Nfields} 
\end{eqnarray} 
 
\noindent  
where all coupling constants are positive: $\lambda$, 
$f_{j},g_{j}^2, h_{k,\phi}, h_{kj,\chi}$ $> 0$.   
 
As before, we are interested in obtaining the EOM 
for a scalar field configuration $\varphi = \langle \Phi \rangle$. 
{}For this, the fields $\chi_j$ and $\psi_k$ are regarded as part of the  
environment. Once again the scalar field 
$\Phi$ is decomposed into its expectation value and 
fluctuation, $\Phi = \varphi + \phi$, 
where $\langle \Phi \rangle = \varphi$.  
The EOM for $\varphi$ then is obtained from 
the tadpole method by imposing 
that $\langle \phi \rangle = 0$, which 
leads to the condition that the sum of all tadpole terms 
vanish. Restricting again our analysis of the EOM to a homogeneous field 
$\varphi \equiv \varphi (t)$, we obtain the effective EOM for $\varphi$ 
 
\begin{eqnarray} 
&&\ddot{\varphi}(t) + m_\phi^2 \varphi(t) + \frac{\lambda}{6} \varphi^3(t)  
+\frac{\lambda}{2} \varphi(t) \langle \phi^2 \rangle 
+\frac{\lambda}{6} \langle \phi^3 \rangle 
+\sum_{j=1}^{N_{\chi}} g_j^2 \left[\varphi (t) \langle \chi_j^2 \rangle + 
\langle \phi \chi_j^2 \rangle \right] \nonumber \\ 
&& + 
\sum_{k=1}^{N_{\psi}} h_{k,\phi} \langle \bar{\psi_k} \psi_k \rangle= 0 \;, 
\label{eq2phi1} 
\end{eqnarray} 
 
\noindent  
where the field averages above  
can be expressed as usual in terms of the coincidence limit of the  
(causal) two-point 
Green's functions $G^{++}_\phi (x,x')$, $G^{++}_{\chi} (x,x')$ 
and $S^{++}_\psi (x,x')$ for the $\Phi$, $\chi_j$ and $\psi_k$ fields, 
respectively. Using the expressions of the previous  
Section for $G^{++}$ and $S^{++}$ for 
the scalar and fermionic propagators respectively, and working out 
the expression analogous to Eq. (\ref{eq1phi4}), we obtain the EOM

\begin{eqnarray} 
&& \ddot{\varphi}(t) + \bar{m}_{\phi}^2 \; \varphi (t) +  
\frac{\bar{\lambda}}{6}  
\varphi^3 (t) \nonumber \\ 
&&+ \lambda^2 \varphi (t) \int_{-\infty}^t dt'\; 
\varphi (t') \dot{\varphi} (t') \int \frac{d^3 q}{(2 \pi)^3} 
\frac{ \left[\omega_\phi \cos(2 \omega_\phi |t-t'|) + \Gamma_\phi  
\sin(2 \omega_\phi 
|t-t'|) \right] }{ 8 \omega_\phi^2 \left( \Gamma_\phi^2 + \omega_\phi^2 
\right) } e^{-2 \Gamma_\phi(q) |t-t'|} \nonumber \\ 
&& + \frac{\lambda^2}{3} \int_{-\infty}^{t} d t' \; 
\dot{\varphi} (t') \int \frac{d^3 {\bf q}_1}{(2 \pi)^3}  
\frac{d^3 {\bf q}_2}{(2 \pi)^3} 
\left\{ \left[\omega_\phi ({\bf q}_1) + \omega_\phi ({\bf q}_2) + 
\omega_\phi ({\bf q}_1+{\bf q}_2)\right] \right. \nonumber \\ 
&& \times \left. \cos[(\omega_\phi ({\bf q}_1) +  
\omega_\phi ({\bf q}_2) 
+\omega_\phi ({\bf q}_1+{\bf q}_2) )|t-t'|\,]  \right.  
\nonumber \\  
&& \left. + \left[\Gamma_\phi ({\bf q}_1) + \Gamma_\phi ({\bf q}_2) + 
\Gamma_\phi ({\bf q}_1+{\bf q}_2)\right] \sin[(\omega_\phi ({\bf q}_1) +  
\omega_\phi ({\bf q}_2) 
+\omega_\phi ({\bf q}_1+{\bf q}_2) )|t-t'|\,]\right\} \nonumber \\ 
&& \times \frac{ 
e^{-[\Gamma_\phi ({\bf q}_1) + \Gamma_\phi ({\bf q}_2) + 
\Gamma_\phi ({\bf q}_1+{\bf q}_2) ] |t-t'|}}{8 \omega_\phi ({\bf q}_1)  
\omega_\phi ({\bf q}_2) \omega_\phi ({\bf q}_1+{\bf q}_2)}  
\nonumber \\ 
&& \times \frac{1}{  \left[\Gamma_\phi ({\bf q}_1) +  
\Gamma_\phi ({\bf q}_2) + 
\Gamma_\phi ({\bf q}_1+{\bf q}_2) \right]^2 + \left[\omega_\phi ({\bf q}_1) + 
\omega_\phi ({\bf q}_2) +\omega_\phi ({\bf q}_1+{\bf q}_2)\right]^2  } 
\nonumber \\ 
& & + \sum_{j=1}^{N_\chi} g_j^4 \;\varphi (t)  
\int_{-\infty}^t  dt' \;\varphi (t') \dot{\varphi} (t')  
\int \frac{d^3 q}{(2 \pi)^3} 
\frac{ \left[ \omega_{\chi_j} \cos(2 \omega_{\chi_j} |t-t'|) 
+ \Gamma_{\chi_j} \sin(2 \omega_{\chi_j} |t-t'|) \right]} 
{2 \omega_{\chi_j}^2 \left( \Gamma_{\chi_j}^2 + \omega_{\chi_j}^2 \right)} 
e^{-2 \Gamma_{\chi_j} |t - t'|} \nonumber \\ 
&& + \sum_{j=1}^{N_\chi} g_j^4 \int_{-\infty}^{t} d t'\;  
\dot{\varphi} (t') \int \frac{d^3 {\bf q}_1}{(2 \pi)^3}  
\frac{d^3 {\bf q}_2}{(2 \pi)^3} 
\left\{ \left[\omega_{\chi_j} ({\bf q}_1) + \omega_\phi ({\bf q}_2) + 
\omega_\phi ({\bf q}_1+{\bf q}_2)\right] \right. \nonumber \\ 
&& \times \left. \cos[(\omega_{\chi_j} ({\bf q}_1) +  
\omega_\phi ({\bf q}_2) 
+\omega_\phi ({\bf q}_1+{\bf q}_2) )|t-t'|\,]  \right.  
\nonumber \\  
&& \left. + \left[\Gamma_{\chi_j} ({\bf q}_1) + \Gamma_\phi ({\bf q}_2) + 
\Gamma_\phi ({\bf q}_1+{\bf q}_2)\right] \sin[(\omega_{\chi_j} ({\bf q}_1) +  
\omega_\phi ({\bf q}_2) 
+\omega_\phi ({\bf q}_1+{\bf q}_2) )|t-t'|\,]\right\} \nonumber \\ 
&& \times \frac{ 
e^{-[\Gamma_{\chi_j} ({\bf q}_1) + \Gamma_\phi ({\bf q}_2) + 
\Gamma_\phi ({\bf q}_1+{\bf q}_2) ] |t-t'|}}{8 \omega_{\chi_j} ({\bf q}_1)  
\omega_\phi ({\bf q}_2) \omega_\phi ({\bf q}_1+{\bf q}_2)}  
\nonumber \\ 
&& \times \frac{1}{  \left[\Gamma_{\chi_j} ({\bf q}_1) +  
\Gamma_\phi ({\bf q}_2) + 
\Gamma_\phi ({\bf q}_1+{\bf q}_2) \right]^2 + \left[\omega_{\chi_j}  
({\bf q}_1) + 
\omega_\phi ({\bf q}_2) +\omega_\phi ({\bf q}_1+{\bf q}_2)\right]^2  } 
\nonumber \\ 
& & + \sum_{k=1}^{N_\psi} h_{k,\phi}^2 \;  
\int_{-\infty}^t  dt' \; \dot{\varphi} (t')  
\int \frac{d^3 q}{(2 \pi)^3} 
\frac{e^{-2 \Gamma_{\psi_k} (q) |t-t'|}}{ \omega_{\psi_k}^2 m_{\psi_k}^2  
\left( \Gamma_{\psi_k}^2 + \omega_{\psi_k}^2 \right) } 
\nonumber \\   
&& \times \left\{ \left[ \Gamma_{\psi_k}^4 \omega_{\psi_k} +  
2 \omega_{\psi_k}^3 
m_{\psi_k}^2 - 2 m_{\psi_k}^4 \omega_{\psi_k} + \Gamma_{\psi_k}^2 
\omega_{\psi_k} (\omega_{\psi_k}^2 + m_{\psi_k}^2) \right] 
\cos(2 \omega_{\psi_k} |t-t'|) \right. \nonumber \\  
&& + \left. \left[ \Gamma_{\psi_k}(2 m_{\psi_k}^2 - \Gamma_{\psi_k}^2)  
(\omega_{\psi_k}^2 - m_{\psi_k}^2) -\Gamma_{\psi_k}^5 \right] 
\sin(2 \omega_{\psi_k}|t-t'|) \right\} 
= 0\;. 
\label{eq2phi2} 
\end{eqnarray}

The dissipative dynamics in these indirect decay models 
will differ based on the relation of the masses 
amongst the $\Phi$, $\chi_j$ and $\psi_i$ fields. 
Consider first the case where the (effective) masses satisfy the  
relation 
 
\begin{equation} 
\label{cascade} 
m_\phi > m_{\chi_j} > 2 m_{\psi_k} \;, 
\end{equation} 
 
\noindent 
which we will refer to more specifically as  
the indirect cascade decay regime.  The first attribute of 
this regime to note is at zero temperature  
$\Gamma_{\phi}$ and $\Gamma_{\chi_j}$ are non-vanishing 
and $\Gamma_\psi=0$. There are two kinematically allowed 
on-shell processes,  
$\Phi \to \psi_k + \bar{\psi}_k$, with decay width $\Gamma_\phi$  
as given in Eq. (\ref{Gamma1}), and  
$\chi_j \to \psi_k + \bar{\psi}_k$, with decay width 
 
\begin{equation} 
\Gamma_{\chi_j} (q) = \sum_{k=1}^{N_\psi}  
\frac{h_{kj,\chi}^2}{8 \pi \omega_{\chi_j}} m_{\chi_j}^2 
\left(1 - \frac{4 m_{\psi_k}^2}{m_{\chi_j}^2} \right)^{\frac{3}{2}}\;. 
\label{gammachi} 
\end{equation} 
Implementing once again the adiabatic-Markovian approximation 
of Subsec. IIA, 
we then obtain the same expression for the EOM as 
Eq. (\ref{final1}), except $\eta(\varphi)$ now is given by 
 
\begin{eqnarray} 
&&\eta(\varphi)  =  
\varphi^2 (t) \frac{\lambda^2 \alpha_{\phi,\psi}^2 }{128 \pi \;  
\sqrt{m_{\phi}^4 +  
\alpha_{\phi,\psi}^4}\; \sqrt{2 \sqrt{m_{\phi}^4 + \alpha_{\phi,\psi}^4} + 
2m_{\phi}^2}}\nonumber \\ 
&& +  \frac{\lambda^2 \alpha_{\phi,\psi}^2}{4}  
\int \frac{d^3 {\bf q}_1}{(2 \pi)^3}  
\frac{d^3 {\bf q}_2}{(2 \pi)^3}  
\left\{ 
\frac{1}{\omega_\phi ({\bf q}_1)^2  
\omega_\phi ({\bf q}_2) \omega_\phi ({\bf q}_1+{\bf q}_2) 
\left[\omega_\phi ({\bf q}_1) + 
\omega_\phi ({\bf q}_2) +\omega_\phi ({\bf q}_1+{\bf q}_2)\right]^3}  
+ {\cal O}\left(\frac{\Gamma_{\phi}^2}{\omega_{\phi}^2} \right) \right\} 
\nonumber \\ 
&& +  \varphi^2(t) \sum_{j=1}^{N_\chi} g_j^4  
\frac{\alpha_{\chi,\psi}^2}{32 \pi} 
\frac{1}{\sqrt{m_{\chi_j}^4 + \alpha_{\chi,\psi}^4}\;  
\sqrt{2 \sqrt{m_{\chi_j}^4 + \alpha_{\chi,\psi}^4} + 
2m_{\chi_j}^2}}  \nonumber \\ 
&& + \sum_{j=1}^{N_\chi} \frac{g_j^4 \alpha_{\chi,\psi}^2}{4} \!\! 
\int \frac{d^3 {\bf q}_1}{(2 \pi)^3}  
\frac{d^3 {\bf q}_2}{(2 \pi)^3}  
\left\{ 
\frac{1}{\omega_{\chi_j} ({\bf q}_1)^2  
\omega_\phi ({\bf q}_2) \omega_\phi ({\bf q}_1+{\bf q}_2) 
\left[\omega_{\chi_j} ({\bf q}_1) + 
\omega_\phi ({\bf q}_2) +\omega_\phi ({\bf q}_1+{\bf q}_2)\right]^3}  
\right. \nonumber \\ 
&& \left. + {\cal O}\left(\frac{\Gamma_{\chi_j}^2,\Gamma_{\phi}^2}{\omega^2} 
\right) \right\}, 
\label{eta2} 
\end{eqnarray} 
 
\noindent 
with $\alpha_{\phi,\psi}$ given by Eq. (\ref{alpha}) and   
 
\begin{equation} 
\alpha_{\chi,\psi}^2 = \sum_{k=1}^{N_\psi} \frac{h_{kj,\chi}^2}{8 \pi}  
m_{\chi_j}^2 
\left(1-\frac{4 m_{\psi_k}^2}{m_{\chi_j}^2} \right)^{\frac{3}{2}} \;. 
\label{alpha2} 
\end{equation} 
 
It is interesting to note that the last two terms in Eq. (\ref{eta2})  
come solely from the decay channels of $\chi_j$ into the fermion 
fields $\psi_k$, which then backreact on the system field  
$\varphi$ in terms of a damping force. The same effect also 
would appear for the case where $\Phi$ was the lightest field,  
 
\begin{equation} 
\label{indirect} 
m_{\chi_j} > 2 m_{\psi_k} > m_\phi. 
\end{equation} 
 
\noindent 
This regime will be referred to simply as the indirect decay regime 
outside the cascade region Eq. (\ref{cascade}). 
{}For this case $\Gamma_\phi=0$, or $\alpha_{\phi,\psi}=0$ in  
Eq. (\ref{eta2}). This type of dissipation, in which the system field 
is lighter than the decay products was 
first noted by Calzetta and Hu in 
\cite{hu4}.  They have shown how a heavy field influences the dynamics 
of a light field in the form of dissipation and fluctuations of the  
light field, even when no aspect of the light field 
dynamics is above the mass threshold of the heavy field. 
This same behavior also can be inferred from our results for the above regime. 
 
Let us finally make a few comments about higher order loop terms. 
Observe that the results obtained for the dissipation  
coefficients in Eqs. (\ref{eta1}) and (\ref{eta2}) are the 
leading order ones at zero temperature, ${\cal O}(\lambda^2 h_k^2)$ and  
${\cal O}(g_j^4 h_{kj,\chi}^2)$ respectively.  
We have neglected higher order loop contributions 
to the EOM, since they all can be shown also to be of higher order  
in the coupling constants. {}For instance, a two-loop contribution made 
of a fermion loop with a vertical scalar  
propagator can 
easily be seen to give a contribution to the dissipation coefficient 
$\eta(\varphi)$ of order  
${\cal O}(h_k^4 \Gamma_\phi) \sim {\cal O}(h_k^6)$, for  
the first case analyzed in Sec. \ref{sect2}, and ${\cal O}(h_{k,\phi}^2 
h_{kj,\chi}^4)$, for the case studied in this section. Also, higher order 
scalar loop terms are subleading in the coupling constants, 
as compared to the results given by Eqs. (\ref{eta1}) and 
(\ref{eta2}). {}For example, at finite temperature, 
scalar ladder diagrams \cite{BGR,jeon}  
are known to be of the same order 
as the two-vertex one-loop term in the EOM, 
due to on-shell divergences of these diagrams and the way  
the field decay widths regularize them.  However at zero temperature  
these divergences are not present and ladder diagrams are  
at most of order ${\cal O}(\lambda^4 \Gamma_\phi^2) \sim 
{\cal O}(\lambda^4 h_k^4)$, for the model in Sec. \ref{sect2}, and 
${\cal O}(g_j^8 h_{kj,\chi}^4)$, for the model in this  
section.

\section{Application} 
\label{sect4} 
 
In this Section we will apply the
effective equation
of motion Eq. (\ref{final1}), with $\eta(\varphi)$  
as given by Eqs. (\ref{eta1}) and 
(\ref{eta2}), to make  estimates of
entropy production from conversion of the scalar field 
potential energy into radiation.   Recall the applicability
of Eq. (\ref{final1}) is limited, since it requires the validity
of the adiabatic-Markovian approximation of Sect. \ref{sect2}.
For this reason and since
as mentioned in the Introduction,
the zero temperature dissipation found here is indicative of
a nonequilibrium dynamics that drives the system to finite temperature,
we will not delve into detailed applications of this effective
equation of motion.
The full dynamics of this problem 
must be understood before detailed application
is worthwhile.  Nevertheless, we will make some naive
estimates of radiation production from our
equations, just to get a feeling for the magnitudes of
the effect. Note that for application to warm inflation, 
the fact dissipation occurs at $T=0$ implies the dynamics 
automatically will bootstrap the universe to finite temperature, 
independent of initial conditions.   
 
In general for a damped equation of motion of the form (\ref{final1}) 
the regime of overdamped (underdamped) motion is 
\begin{equation} 
m^2(\varphi) = m_{\phi}^2 + \frac{\lambda}{2} \varphi^2 \; 
< (>) \; \eta^2(\varphi) . 
\label{odud} 
\end{equation} 
{}For the direct decay models in Sec. \ref{sect2},  
since the dissipative 
coefficient $\eta(\varphi)$ is suppressed by two powers of the $\phi$  
self-coupling parameter $\lambda$, we find  
that underdamped motion generally is possible, 
unless there are very many environment bath fields. 
{}For the indirect decay models of Sect. \ref{sect3}, 
the cascade decay regime Eq. (\ref{cascade}) has similar constraints, 
thus solutions, as those of the direct decay models of Sect. \ref{sect2}. 
However, in the other interesting regime Eq. (\ref{indirect}), 
the dissipative coefficient and the parameters of the $\varphi$ 
potential can be independently tuned, which 
means the overdamped regime 
can be identified for as few as one or two heat bath 
fields.  Both the underdamped and overdamped regimes 
may have application to the chiral transition in heavy ion collision 
\cite{heavyion}, whereas the overdamped regime also is of interest to 
the warm inflation scenario \cite{wi}.  Below, both the 
direct and indirect decay models will be examined. 
 
{}For the direct decay model, the overdamped (underdamped) 
regime from 
Eqs. (\ref{eta1}), (\ref{alpha}) and (\ref{odud}) is given by 
\begin{equation} 
\frac{\lambda^2 N_{\psi}}{2048 \pi^2} 
\gtrsim (\lesssim) 1, 
\end{equation} 
where the Yukawa coupling $h$ is determined by requiring 
$m_{\psi} \approx h \varphi < m(\varphi)/2  
\approx \sqrt{\lambda} \varphi/2$. 
Thus, even for strong coupling $\lambda \sim 1$, it requires 
$N_{\psi} \gtrsim 10^4$ fields for the overdamped regime. 
 
Considering first the underdamped regime, the adiabatic  
condition Eq. (\ref{kinetic}) requires  
\begin{equation} 
\frac{\dot \varphi}{\varphi} \approx m(\varphi) < \Gamma_{\psi} 
\approx \frac{h^2 N_{\psi}}{8 \pi} m(\varphi) 
\approx \frac{\lambda N_{\psi}}{16 \pi} m(\varphi), 
\label{ddud} 
\end{equation} 
where for underdamped motion from Eq. (\ref{final1})  
${\dot \varphi} \lesssim m(\varphi) \varphi$. 
The energy dissipated by the scalar field goes into  
radiation energy density $\rho_r$, 
here composed of fermions and/or scalar bosons, at the rate   
\begin{equation} 
{\dot \rho}_r = -\frac{dE_{\phi}}{dt} = 
\eta(\varphi) {\dot \varphi}^2 \approx  
\frac{\lambda^2 N_{\psi} m^5(\varphi)}{2048 \pi^2}, 
\end{equation} 
where to obtain the last expression we estimate 
${\dot \varphi}^2 \approx m^2(\varphi) \varphi^2$. 
{}For the overdamped regime from Eq. (\ref{final1}) 
${\dot \varphi} = m^2(\varphi) \varphi/\eta(\varphi)$.  Writing  
$\eta(\varphi)= Q m(\varphi)$, from Eqs. (\ref{eta1}) and (\ref{alpha}) 
$Q \approx \lambda^2 N/(2048 \pi^2)$ and for overdamping 
Eq. (\ref{odud}) requires $Q>1$.   This requirement 
of overdamping automatically 
implies the adiabatic condition Eq. (\ref{kinetic}) is satisfied.  
Thus the radiation production in this case is 
\begin{equation} 
{\dot \rho}_r \approx 
\frac{m^3(\varphi) \varphi^2}{Q} . 
\label{ddod} 
\end{equation} 
 
Overall, for the direct decay model, the underdamped 
regime is generic except if there are a very large number of bath fields 
$N > 10^4$, in which case overdamped motion also becomes possible. 
In the underdamped regime, moderate radiation energy production 
occurs.  In particular during a characteristic oscillation 
time $\sim 1/m(\varphi)$, the produced radiation has an 
associated temperature scale  
$T \approx (\rho_r/N_{\psi})^{1/4} \approx \lambda^{1/2} m(\varphi)/10 
\lesssim m(\varphi)$.  On the other hand, the overdamped 
regime, although requiring a large number of bath fields, 
can yield sizable radiation by increasing the amplitude 
$\varphi$ in Eq. (\ref{ddod})  
 
Turning next to the indirect decay models of Sec. \ref{sect3}, 
the cascade region Eq. (\ref{cascade}) leads to similar solutions 
as given above for the direct decay models, 
so will not be further elaborated. 
However, dissipative dynamics also appears to occur 
in the regime Eq. (\ref{indirect}), which does not 
have a direct interpretation in terms of particle decay 
at one of the two steps of the process, the 
$\varphi \rightarrow \chi$ transition.  We believe further investigation 
is needed of this case in order to obtain a sensible interpretation 
of this process, and this is left 
for future work.  Nevertheless, here it is interesting 
to estimate the size of the dissipative effects for 
this case.  In particular for this case, the overdamped 
regime can easily be obtained as will be shown next. 
The overdamped regime requires the condition in Eq. (\ref{odud}) 
and the adiabatic condition requires   
\begin{equation} 
\frac{\dot \varphi}{\varphi} = \frac{m^2(\varphi)}{\eta(\varphi)} <  
\Gamma_{\chi}. 
\label{adcond} 
\end{equation} 
Since the scalar field sector has two free parameters, 
$m_{\phi}$ and $\lambda$, it is always possible to tune $m^2(\varphi)$ 
to satisfy both the above requirements independent of 
$\Gamma_{\chi}$, $\eta(\varphi)$, and for any amplitude 
$\varphi$.   
 
In the overdamped regime, the kinetic energy of the scalar field 
is negligible.  Thus, the loss in its potential energy 
translates into the energy released into radiation 
$\rho_r$ as 
\begin{eqnarray} 
{\dot \rho}_r(t) & = & \eta(\varphi) {\dot \phi}^2 
= - \frac{dV}{d \varphi} {\dot \varphi} 
\nonumber\\ 
& = & (m_{\phi}^2 \varphi + \frac{\lambda}{6} \varphi^3 ) 
{\dot \varphi} 
\approx V(\varphi) \frac{m^2(\varphi)}{\eta}  
< V(\varphi) \Gamma_{\chi}, 
\end{eqnarray} 
where the second line follows from Eqs. (\ref{final1}) 
and (\ref{Nfields}). 
Generally the microscopic scale is determined by 
$\Gamma_{\chi}$. In this time interval, we find the radiation 
to increase to 
\begin{equation} 
\label{rhogam} 
\rho_r(1/\Gamma_{\chi}) \approx V(\varphi)  
\frac{m^2(\varphi)}{\eta \Gamma_{\chi}} < V(\varphi), 
\end{equation} 
where to obtain the right most expression 
we used Eq. (\ref{adcond}). 
Thus the energy dissipated into radiation is proportional 
to the potential energy contained in the scalar field. 
To consider some numbers, for example typical for inflation, suppose 
the potential energy is at the GUT scale 
$V(\varphi)^{1/4} \sim 10^{15-16} {\rm GeV}$ 
and $m^2(\varphi)/(\eta \Gamma_{\chi}) \approx 10^{-4}$. 
{}For this, it implies 
a radiation component is generated which, if expressed 
in terms of temperature, is at the scale $T \sim 10^{14-15} {\rm GeV}$, 
and this is nonnegligible. 
The main point to note is that considerable radiation 
production can occur from the indirect decay models 
in the regime Eq. (\ref{indirect}).  
 
\section{Conclusion} 
\label{sect5} 
 
Although the calculations in this paper were for non-expanding Minkowski 
spacetime, a brief reflection will be made here on the 
consequences of these results for inflationary 
cosmology.  In this case, the background 
space is expanding.  The effect of the interaction 
between the scalar field, in this case called the inflaton, 
and the background metric is known to yield a  
$3H{\dot \varphi}$ term in the inflaton effective equation 
of motion, where the Hubble parameter $H \equiv {\dot a}/a$ 
and $a(t)$ is the cosmic scale factor.  As well known, 
this term does not arise from microscopic interactions with 
other fields, but rather from the macroscopic interaction with the 
background metric of gravity.  Precisely this disparity 
in scales and the difference in origin of the interactions 
suggests that the effect of this term and the dissipative term 
computed in this paper will act independently on the inflaton 
to a good approximation, with perhaps some 
self-consistency requirements.  At present, we are extending 
our calculation to expanding spacetime in order to 
examine this point.  Should this work render true 
the expectations from Sec. \ref{sect4}  
for dissipation, it will be difficult to 
justify the supercooled inflation picture and rather it would 
appear the warm inflation picture is the natural one. 
 
{}For the moment, assuming the correctness of the above 
expectations, a lower bound on the temperature of the universe 
during inflation can be estimated from the direct decay 
model.  In typical inflation models,  
$m_{\phi} \sim H$ for supercooled inflation and  
$m_{\phi}$ perhaps a few 
orders of magnitude bigger than $H$ for warm inflation. 
{}Furthermore, the conditions on density perturbations 
generally require the inflaton self-coupling 
parameter to be tiny $\lambda \sim 10^{-(10 - 16)}$. 
In an expanding background, the evolution of the radiation 
in presence of a dissipating scalar field source is 
given by Eq. (\ref{ddud}) with the addition of the term $-4H \rho_r$ 
to the right hand side, which accounts for the redshift of 
the radiation due to background expansion. 
Therefore, in steady state ${\dot \rho_r}=0$, 
$\rho_r \approx \eta {\dot \varphi}^2/(4H)$, 
or the associated temperature 
$T \sim (\rho_r/g^*)^{1/4}$, where we will 
take the number of light particles $g^* \sim N$. 
With these estimates, based on Eq. (\ref{ddud}) we find 
$T \gtrsim \lambda^{1/2} m_{\phi}/10$. 
If $m_{\phi} \sim 10^{10} {\rm GeV}$ as a typical 
value, then this implies 
$T \sim (10^4 - 10) {\rm GeV}$. 
 
In order not to affect the successful predictions of  
nucleosynthesis, the primary requirement is that the universe 
should be well within the hot Big-Bang radiation dominated 
regime by $T \approx 10 {\rm MeV}$. Slightly more 
conservative, though not necessarily mandatory, 
is to require that the QCD phase transition at  
$T \lesssim 1 {\rm GeV}$ occurs within the 
radiation dominated era.  So a safe lower bound 
for inflation to end and the radiation dominated era to commence 
is $T \gtrsim 1 {\rm GeV}$.  As such, the lower 
limits for radiation production during inflation given above 
still would be above this lower limit requirement set by cosmology. 
The results found here may be also useful in applications 
to the low temperature regimes of warm inflation identified 
in the phenomenological studies of \cite{radiation}. 
 
It should be clarified that the results found in this paper in no 
way require supersymmetry.  However, these calculations 
easily could be applied in SUSY models.  In such models, 
it is becoming appreciated that to avoid gravitino overproduction, 
for any type of inflation scenario 
the temperature of the universe after inflation can not be 
very high $T \lesssim 10^{10} {\rm GeV}$ \cite{gravitino}. 
{}From this perspective, the possibility found in this paper 
for low temperature warm inflation solutions  
in the direct decay models would be phenomenologically 
attractive. 
 
One cautionary remark is in order.  Since  
$m_{\phi} \sim H$, the lower limit on the temperature 
during inflation that is suggested above, is below the 
so called Gibbons-Hawking "temperature"  
$T_{GH} = H/(2\pi)$.  However $T_{GH}$ does not represent a  
temperature in the usual sense of a thermal bath of particles. 
$T_{GH}$ acts like a temperature in the formal sense that for 
a non-interacting scalar field in de Sitter space, its 
euclideanized de Sitter invariant Green's function 
is periodic in imaginary time.  The role that 
$T_{GH}$ plays in this Green's function is formally the same 
as what actual temperature plays in the static thermal 
Green's function in Minkowski space. 
However, for the non-interacting scalar field in 
the de Sitter case, there are  no particles present in the sense  
that the field is in a vacuum state\footnote{We thank Larry Ford for 
this clarification}.  In contrast, the radiation production we have 
computed for the interacting scalar field results in real particle 
production, irrespective whether its associated temperature 
is above or below $T_{GH}$. 
 
In summary, this paper has studied dissipative effects of interacting scalar 
fields at zero temperature. Similar treatments along this line are limited  
\cite{boya,hu,hu4} and one of the novel features of this paper is the 
appropriate inclusion of quasi-particles effects through the fully dressed  
zero temperature two-point Green's functions.  Another feature 
of our analysis which has been studied only to a limited 
extent in the literature is a 
detailed examination 
of the dissipative kernel in Subsec. \ref{sect2B}. 
The models examined in this paper were generic and in all cases  
dissipation was found. 
Since dissipative effects are seen for the zero temperature state,   
we conclude that radiation production from dissipation 
is invariably present for generic interacting scalar field 
systems, although the extent of radiation production can vary immensely. 
Minimally, it appears the mass of the scalar field times a suitable 
dimensionless coupling constant sets a lower limit to the associated 
temperature scale of the produced radiation. 
However, for the indirect decay models in the region Eq. (\ref{indirect}), 
there appears a much more robust 
possibility for producing radiation. Although formally 
this is what is indicated by our calculations, 
as mentioned earlier, we feel further investigation is necessary 
of these indirect decay models in order to obtain  
a sensible interpretation of its dissipative 
process.  In regards the potential implications of 
the results found in this paper to 
inflationary cosmology, 
we infer that under generic circumstances the scalar inflaton field 
will dissipate a nonnegligible amount of radiation during 
inflation.  In particular,  
the lower bound suggested by the above estimates  
for the direct decay models already are sufficiently 
high to preclude a mandatory requirement for a reheating. 
{}Furthermore, the upper bound from the indirect decay models 
in the regime Eq. (\ref{indirect}) could yield very 
high temperature warm inflation solutions, in the range discussed 
below Eq. (\ref{rhogam}). 
However, these only are expectations suggested by 
the calculations in this paper. Verification of these 
expectations requires a proper extension of these 
calculations to expanding spacetime, which we currently 
are examining.

\acknowledgments 
 
We thank Larry Ford for helpful discussions. 
We especially thank Ian Lawrie for several discussions and 
for a careful examination of our paper. 
AB was funded by the United Kingdom Particle Physics and 
Astronomy Research Council (PPARC) and 
ROR was supported by Conselho Nacional de Desenvolvimento 
Cient\'{\i}fico e Tecnol\'ogico (CNPq-Brazil) and SR2-UERJ.

\appendix 
 
\section{}

In this appendix the decay widths for the  
processes $\phi \to \psi +\bar{\psi}$ (or  
$\chi \to \psi + \bar{\psi}$) and  $\phi \to \chi +\chi$ 
are derived. 
Recall the basic expression for the decay of an  
initial particle of momentum $p$ into two particles, 
 
\begin{eqnarray} 
\Gamma(p) & = & \frac{1}{2\omega_{\bf p}} 
\left[\int \frac{d^4 k_1}{(2\pi)^3} \frac{d^4 k_2}{(2\pi)^3} 
(2 \pi)^4 \delta^{(4)}(k_1+k_2-p) \right.  
\nonumber\\ 
& &  \times \left. \delta(k_1^2 -m_1^2) 
\delta(k_2^2-m_2) \prod_{{\rm fermion}_j} (2m_j) \sum_{{\rm spin} s_j} 
|{\cal M}_{fi}|^2 \right], 
\label{decaywid} 
\end{eqnarray}

\noindent 
where $\omega_{\bf p} = \sqrt{{\bf p}^2 + M^2}$, 
$M$ is the mass of the scalar decaying  
field, and $m$ stands for the mass of the decay products. 
Observe that the expression in the square brackets is 
Lorentz invariant, thus most conveniently it is evaluated 
in the rest frame of the initial particle ${\bf p}=0$. 
 
{}For the scalar to 2 fermion model ($h$ is the Yukawa coupling) 
\begin{equation} 
{\cal L}_I = - h \phi(x) {\bar \psi}(x) \psi(x)\;, 
\end{equation} 
which implies 
\begin{equation} 
{\cal M} = -i h {\bar u}_{s_1}({\bf k}_1) v_{s_2}({\bf k}_2), 
\end{equation} 
so that 
\begin{equation} 
\sum_{s_j} |{\cal M}|^2 =  h^2 \frac{k_1 \cdot k_2 -m^2}{m^2}. 
\end{equation} 
Substituting this into Eq. (\ref{decaywid}) gives 
\begin{equation} 
\Gamma(p) = \frac{h^2M^2}{8\pi \omega (\bf p)}   
\left(1-\frac{4m^2}{M^2}\right)^{3/2}. 
\end{equation} 
 
{}For the scalar to two scalar model, with coupling constant $g$, 
we have instead that 
\begin{equation} 
{\cal L}_I = - \frac{g^2}{2} \phi(x) \chi^2(x), 
\end{equation} 
which implies 
\begin{equation} 
{\cal M} = -i g^2, 
\end{equation} 
and so 
\begin{equation} 
\Gamma(p) = \frac{g^4}{16\pi \omega_{\phi}({\bf p})}   
\left(1-\frac{4m^2}{M^2}\right)^{1/2}. 
\end{equation}

\newpage 
 
\centerline{\bf \large Figure Caption} 
 
{\bf Fig. 1.} The kernel $K(t,0)/\lambda^2$ in 
Eq. (\ref{kernel}) for the cases  
$\Gamma_{\phi}(0)/m_{\phi} = 0.1$ (solid), $0.5$ (dashed), 
$1.0$ (dotted), and $5.0$ (dot-dashed), plotted over 
four different time intervals. 
  
\newpage

\begin{figure}[b] 
\epsfysize=18cm  
{\centerline{\epsfbox{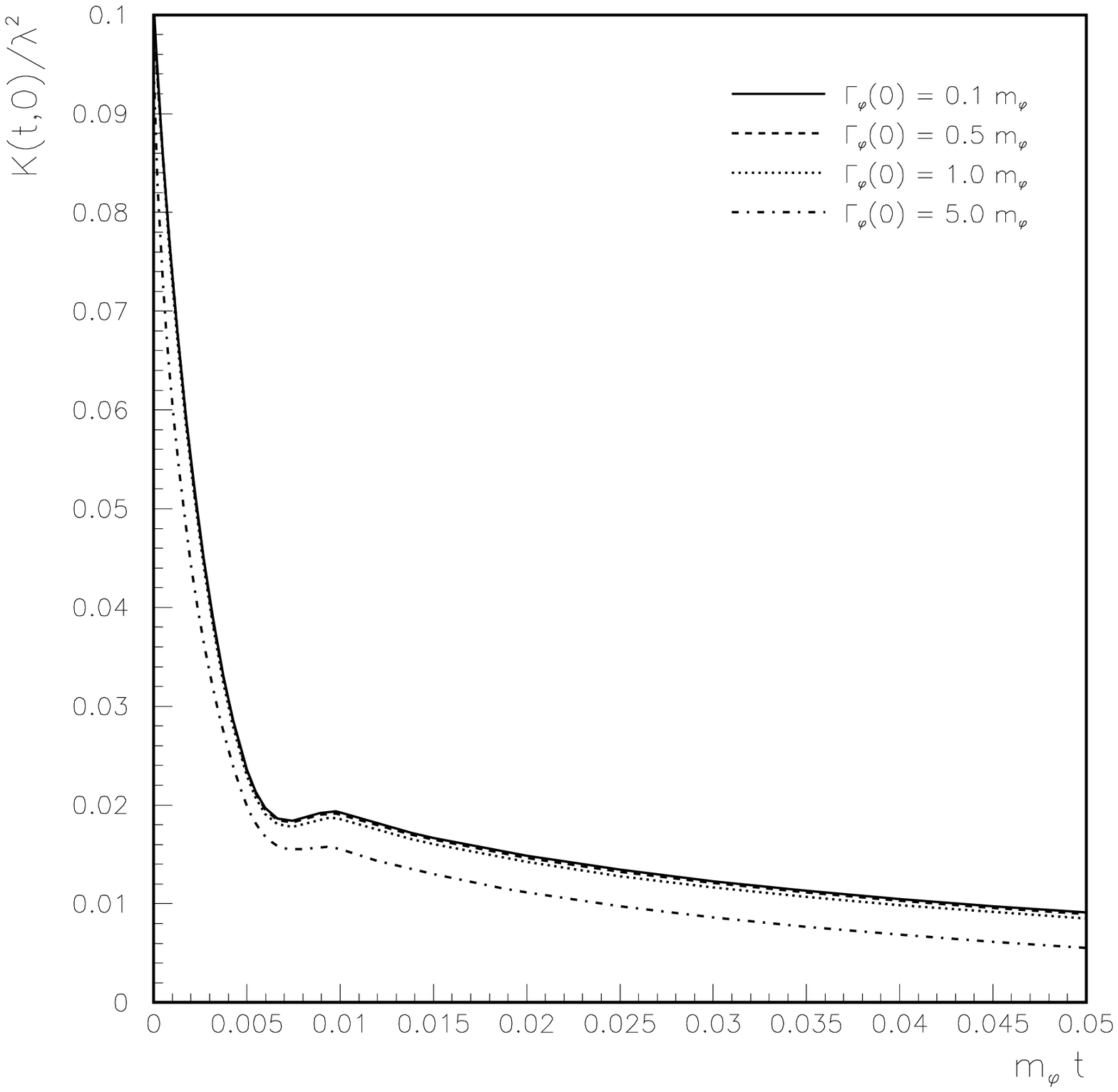}}} 
 
\vspace{1cm} 
 
\end{figure} 
 
\begin{figure}[b] 
\epsfysize=18cm  
{\centerline{\epsfbox{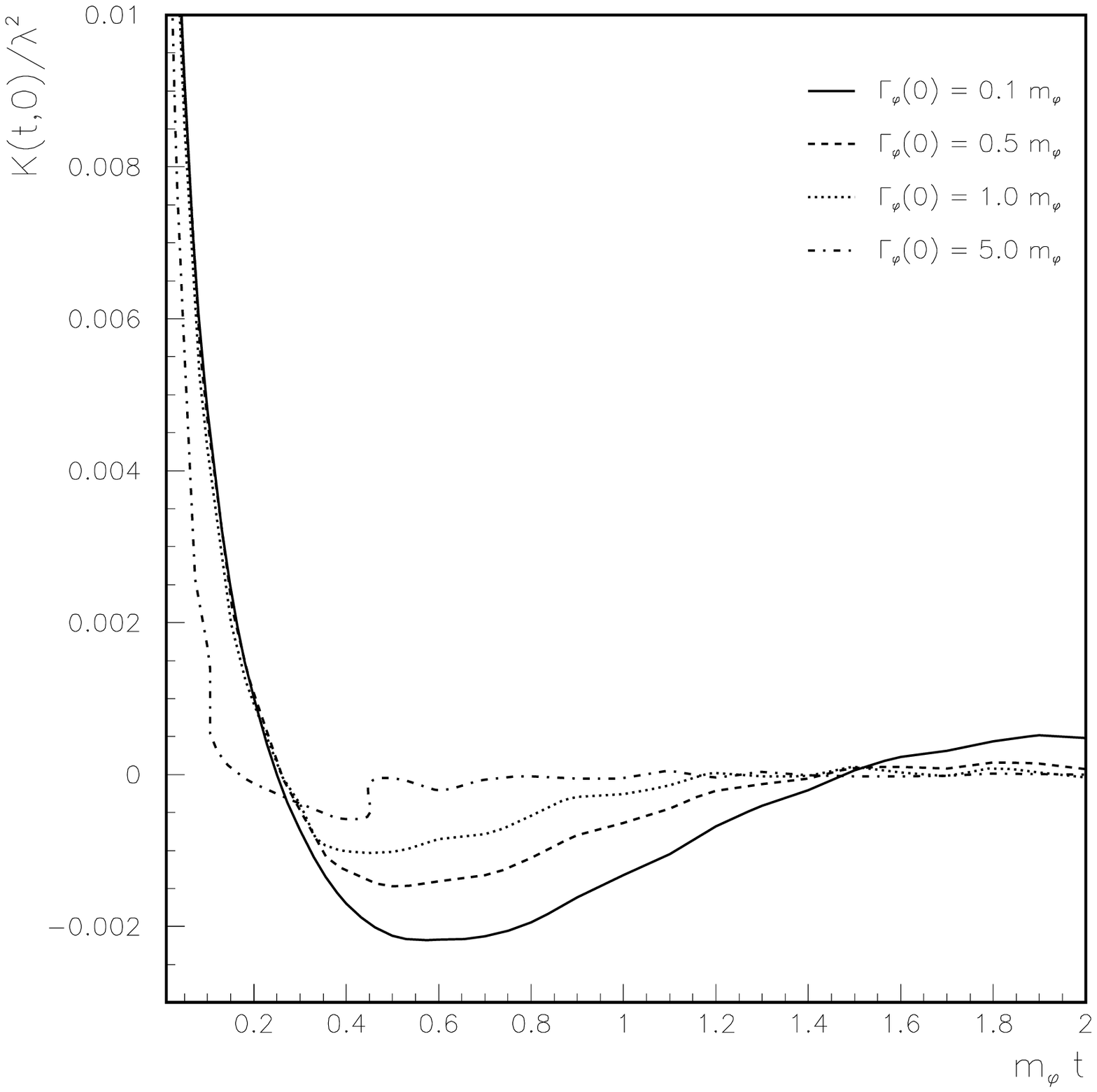}}} 
 
\vspace{1cm} 
 
\end{figure} 
 
\begin{figure}[b] 
\epsfysize=18cm  
{\centerline{\epsfbox{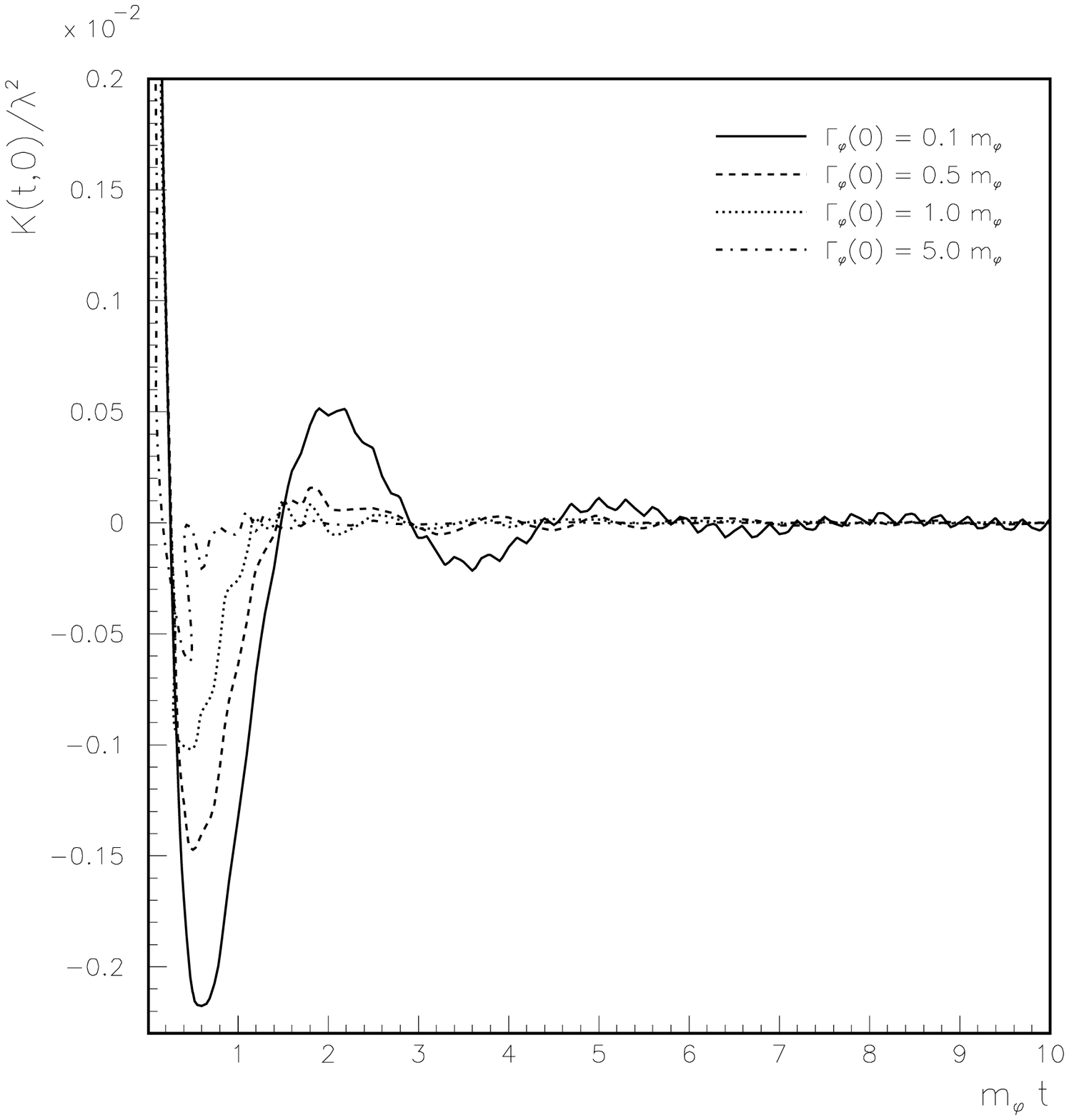}}} 
 
\vspace{1cm} 
 
\end{figure} 
 
\begin{figure}[b] 
\epsfysize=18cm  
{\centerline{\epsfbox{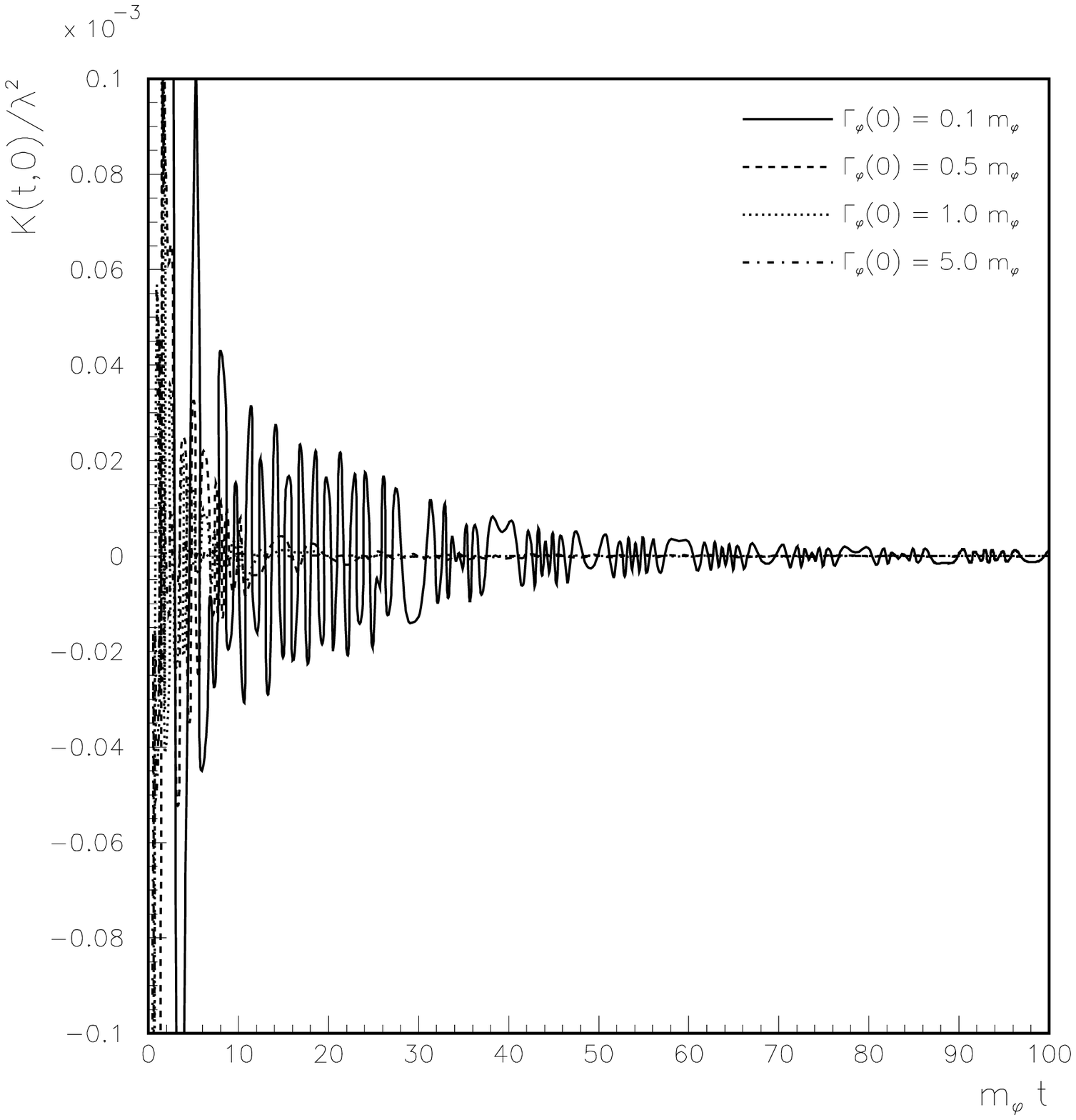}}} 
 
\vspace{1cm} 
 
\end{figure} 
 

\begin{references} 
 
\bibitem{rev} D. Boyanovsky and H. J. de Vega, 
hep-ph/9909372. 
 
 
\bibitem{boya} D. Boyanovsky, H. J. de Vega,  
R. Holman, D. S-Lee and A. Singh, Phys. Rev. {\bf D51}, 
4419 (1995).  
 
\bibitem{boya1} D. Boyanovsky, M. D'attanasio, H.J. de Vega, R. Holman and  
D.-S.Lee, Phys. Rev. {\bf D52},  6805 (1995). 
 
\bibitem{boya3} D. Boyanovsky, H. J. de Vega, R. Holman and J. F. J. Salgado, 
Phys. Rev. {\bf D54}, 7570 (1996); Phys. Rev. {\bf D59}, 
125009 (1999). 
 
 
\bibitem{habib}F. Cooper, S. Habib, Y. Kluger and E. Mottola, 
Phys. Rev. {\bf D55}, 6471 (1997). 
 
\bibitem{hu3} S. A. Ramsey and  B. L. Hu, 
Phys. Rev. {\bf D56},  661 (1997). 
 
\bibitem{baacke2} J. Baacke, K. Heitmann and C. Patzold, 
Phys. Rev. {\bf D57},  6406 (1998). 
 
\bibitem{hs}A. Hosoya and M. Sakagami, Phys. Rev. {\bf D29}, 
2228 (1984). 
 
\bibitem{morikawa}M. Morikawa, Phys. Rev. {\bf D33}, 3607 (1986). 
M. Morikawa and M. Sasaki, Phys. Lett. {\bf 165B}, 59 (1985). 
 
\bibitem{ian}I. D. Lawrie, J. Phys. {\bf A25}, 6493 (1992); 
Phys. Rev. {\bf D60}, 063510 (1999); 
 
\bibitem{GR} M. Gleiser and R. O. Ramos, Phys. Rev. {\bf D50}, 2441 
(1994). 
 
\bibitem{BGR} A. Berera, M. Gleiser and R. O. Ramos, Phys. Rev. {\bf D58}, 
123508 (1998).  
 
\bibitem{hu}E. Calzetta and B. L. Hu, Phys. Rev. {\bf D61}, 025012 (2000); 
Phys. Rev. {\bf D40}, 656 (1989).  
 
\bibitem{hu4}E. Calzetta and B. L. Hu, Phys. Rev. {\bf D55}, 3536 (1997). 
 
\bibitem{patkos}Sz. Borsanyi, A. Patkos, J. Polonyi and Zs. Szep, 
Phys. Rev. {\bf D62}, 085013 (2000).  
 
\bibitem{wett}  G. Aarts, G. F. Bonini and C. Wetterich, 
Nucl. Phys. {\bf B587}, 403 (2000). 
 
\bibitem{cl} A. O. Caldeira and A. J. Leggett, Ann. Phys. {\bf 149}, 
374 (1983). 
 
\bibitem{BGR2} A. Berera, M. Gleiser and R. O. Ramos, Phys. Rev. Lett. 
{\bf 83}, 264 (1999).  
 
\bibitem{abadiab} A. Berera, Nucl. Phys. {\bf B585}, 
666 (2000). 
 
\bibitem{linde}J. Yokoyama and  A. Linde, 
Phys. Rev. {\bf D60},  083509 (1999). 
 
\bibitem{wi} A. Berera,  Phys. Rev. Lett. {\bf 75}, 
3218 (1995); 
Phys. Rev. {\bf D54}, 2519 (1996); 
Phys.\ Rev.\  {\bf D55}, 3346 (1997). 
 
\bibitem{ringwald}G. Semenoff and N. Weiss, Phys. Rev. {\bf D31}, 699 (1985); 
A. Ringwald, Phys. Rev. 
{\bf D36}, 2598 (1987); Ann. Phys. {\bf 177}, 129 (1987); 
Z. Phys. {\bf C34}, 481 (1987). 
   
\bibitem{ian2}I. D. Lawrie and D. B. McKernan 
Phys. Rev. {\bf D62}, 105032 (2000). 
 
\bibitem{greiner} C. Greiner and B. M\"uller, Phys. Rev. {\bf D55}, 
1026 (1997). 
 
\bibitem{RF} R. O. Ramos and F. A. R. Navarro, Phys. Rev. {\bf D62}, 
085016 (2000).  
 
\bibitem{hu2}S. A. Ramsey, B. L. Hu and A. M. Stylianopoulos, 
Phys. Rev. {\bf D57}, 6003 (1998). 
 
\bibitem{baacke} J. Baacke, K. Heitmann and C. Patzold, 
Phys. Rev. {\bf D58}, 125013 (1998). 
 
\bibitem{boya2}J. Baacke, D. Boyanovsky and H. J. de Vega, 
hep-ph/9907337. 
 
 
\bibitem{jeon}S. Jeon, Phys. Rev. {\bf D47}, 4586 (1993); 
Phys. Rev. {\bf D52}, 3591 (1995). 
 
 
\bibitem{heavyion}  for a review of the chiral phase transition 
please see K. Rajagopal, in {\it Quark-Gluon Plasma II}, ed. R. Hwa, 
World Scientific (1995), [hep-ph/9504310];    
see also Z. Xu and C. Greiner, Phys Rev. {\bf D62}, 036012 (2000). 
 
\bibitem{radiation}H. P. de Oliveira and  R. O. Ramos, 
Phys. Rev. {\bf D57}, 741 (1998); 
W. Lee and  L.-Z. Fang, Phys. Rev. {\bf D59}, 083503 (1999). 
 
\bibitem{gravitino} M. Kawasaki and T. Moroi, 
Prog. Theor. Phys. {\bf 93}, 879 (1995). 
 
 
\end{references}
\end{document}